\documentclass[sigchi]{acmart}

\usepackage{xcolor}
\usepackage{xspace}

\newcommand{\add}[1]{#1}
\newcommand{\delete}[1]{}

\definecolor{LeftPanelHL}{HTML}{c45100}    
\definecolor{MiddlePanelHL}{HTML}{1e5631}  
\definecolor{RightPanelHL}{HTML}{4a235a}   
\definecolor{PaperTrail}{HTML}{b35900}     
\definecolor{Baseline}{HTML}{666666}       
\definecolor{LowCostText}{HTML}{0d3d47}    
\definecolor{MediumCostText}{HTML}{806000} 
\definecolor{HighCostText}{HTML}{660000}   

\newcommand{\LowCost}[1]{\textcolor{LowCostText}{\textbf{#1}}\xspace}
\newcommand{\MediumCost}[1]{\textcolor{MediumCostText}{\textbf{#1}}\xspace}
\newcommand{\HighCost}[1]{\textcolor{HighCostText}{\textbf{#1}}\xspace}

\newcommand{\system}{\textsc{PaperTrail}\xspace}

\newcommand{\LeftHL}[1]{\textcolor{LeftPanelHL}{\textbf{#1}}}
\newcommand{\MiddleHL}[1]{\textcolor{MiddlePanelHL}{\textbf{#1}}}
\newcommand{\RightHL}[1]{\textcolor{RightPanelHL}{\textbf{#1}}}

\newcommand{\LeftPanel}{\LeftHL{Left Panel (A)}\xspace}
\newcommand{\TaskContext}{\LeftHL{Task Context (A1)}\xspace}
\newcommand{\ReferencesList}{\LeftHL{References List (A2)}\xspace}
\newcommand{\TextEditor}{\LeftHL{Text Editor (A3)}\xspace}
\newcommand{\MiddlePanel}{\MiddleHL{Middle Panel (B)}\xspace}
\newcommand{\ChatInterface}{\MiddleHL{Chat Interface (B1)}\xspace}
\newcommand{\QuestionBank}{\MiddleHL{Question Bank (B2)}\xspace}
\newcommand{\ChatControls}{\MiddleHL{Chat Controls (B3)}\xspace}

\newcommand{\AnswerClaims}{\MiddleHL{Answer Claims (B4)}\xspace}
\newcommand{\SourceHighlight}{\MiddleHL{Source Highlight (B5)}\xspace}
\newcommand{\SourceHighlights}{\MiddleHL{Source Highlights (B5)}\xspace}
\newcommand{\RightPanel}{\RightHL{Right Panel (C)}\xspace}
\newcommand{\ClaimCoverage}{\RightHL{Claim Coverage (C1)}\xspace}
\newcommand{\PaperClaim}{\RightHL{Paper Claim (C2)}\xspace}
\newcommand{\PaperClaims}{\RightHL{Paper Claims (C2)}\xspace}
\newcommand{\PaperSource}{\RightHL{Paper Source (C3)}\xspace}

\AtBeginDocument{%
  }

\copyrightyear{2026}
\acmYear{2026}
\setcopyright{usgovmixed}
\acmConference[CHI '26]{Proceedings of the 2026 CHI Conference on Human Factors in Computing Systems}{April 13--17, 2026}{Barcelona, Spain}
\acmBooktitle{Proceedings of the 2026 CHI Conference on Human Factors in Computing Systems (CHI '26), April 13--17, 2026, Barcelona, Spain}
\acmPrice{}
\acmDOI{10.1145/3772318.3791101}
\acmISBN{979-8-4007-2278-3/2026/04}

\begin{document}

\title[An Interface for Grounding Provenance in LLM-based Scholarly Q\&A]{\system: A Claim-Evidence Interface for \\Grounding Provenance in LLM-based Scholarly Q\&A}

\author{Anna Martin-Boyle}
\email{mart5877@umn.edu}
\affiliation{%
  \institution{University of Minnesota}
  \city{Minneapolis}
  \state{Minnesota}
  \country{USA}
}

\author{Cara A.C. Leckey}
\email{cara.ac.leckey@nasa.gov}
\affiliation{%
  \institution{NASA Langley Research Center}
  \city{Hampton}
  \state{Virginia}
  \country{USA}
}
\author{Martha C. Brown}
\email{martha.c.brown@nasa.gov}
\affiliation{%
  \institution{NASA Langley Research Center}
  \city{Hampton}
  \state{Virginia}
  \country{USA}
}

\author{Harmanpreet Kaur}
\email{harmank@umn.edu}
\affiliation{%
  \institution{University of Minnesota}
  \city{Minneapolis}
  \state{Minnesota}
  \country{USA}
}

\renewcommand{\shortauthors}{Martin-Boyle et al.}

\begin{abstract}

  Large language models (LLMs) are increasingly used in scholarly question-answering (QA) systems to help researchers synthesize vast amounts of literature. However, these systems often produce subtle errors (e.g., unsupported claims, errors of omission), and current provenance mechanisms like source citations are not granular enough for the rigorous verification that scholarly domain requires. To address this, we introduce \system, a novel interface that decomposes both LLM answers and source documents into discrete claims and evidence, mapping them to reveal supported assertions, unsupported claims, and information omitted from the source texts. We evaluated \system in a within-subjects study with 26 researchers who performed two scholarly editing tasks using \system and a baseline interface. Our results show that \system significantly lowered participants' trust compared to the baseline. However, this increased caution did not translate to behavioral changes, as people continued to rely on LLM-generated scholarly edits to avoid a cognitively burdensome task. We discuss the value of claim-evidence matching for understanding LLM trustworthiness in scholarly settings, and present design implications for cognition-friendly communication of provenance information. 
\end{abstract}






\begin{teaserfigure}
    \centering
    \includegraphics[width=\textwidth]{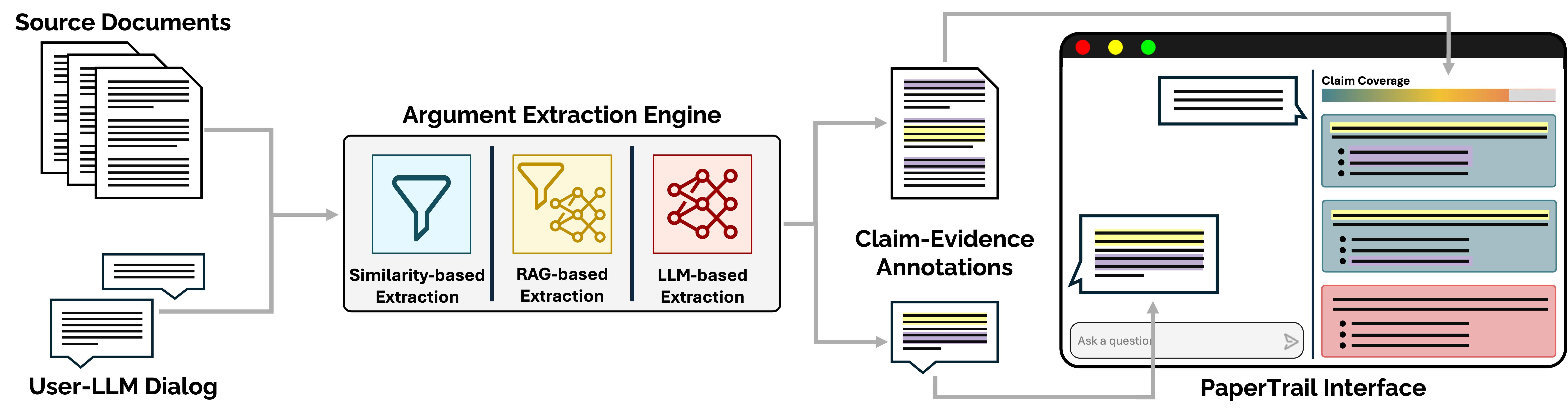}
    \Description{A system architecture diagram showing PaperTrail's three-stage pipeline for argument-grounded provenance. On the left, two inputs feed into the system: source document PDFs (represented as stacked document icons) and user-LLM dialog (shown as a chat interface). These flow into a central box labeled ``Argument Extraction Engine,'' which contains three extraction method icons arranged left to right: similarity-based extraction, RAG-based extraction, and LLM-based extraction. The engine outputs "Claim-Evidence Annotations," depicted as annotated document cards, which feed into the PaperTrail Interface on the right. The interface is shown as a multi-panel application with a ``Claim Coverage'' indicator bar in the upper right corner and a question input field at the bottom.}
    \caption{\system system architecture showing the three-stage pipeline for argument-grounded provenance. Paper PDFs and user-LLM dialog feed into the Argument Extraction Engine, which \add{combines three different extraction methods deployed strategically across pipeline stages based on design-time tradeoffs between computational cost and semantic capability.} The system produces claim-evidence annotations that power the \system interface for scholarly question-answering.}
    \vspace{0.75pc}
    \label{fig:placeholder}
\end{teaserfigure}

\maketitle
\section{Introduction}

The accelerating growth rate of scientific literature presents exciting opportunities for advancing knowledge, yet creates significant challenges as domain experts face escalating cognitive demands in monitoring and synthesizing knowledge within their fields \cite{Bornmann2021,doi:10.1080/0194262X.2021.1991546,doi:10.1080/0194262X.2020.1758284,doi:10.1080/0194262X.2018.1445063,Landhuis2016}. Large language models (LLMs) have emerged as promising solutions for this information overload crisis \cite{Doneva2024,Ineichen2023,Krenn2023}. LLMs are leveraged by end-users for a variety of scholarly tasks \cite{10.1145/3711000,liao2024llmsresearchtoolslarge}, and are being integrated into scholarly question-answering (QA) systems like Semantic Scholar's ``Ask This Paper''\cite{s2qa}, JSTOR's AI research tool \cite{jstor}, and Elicit AI \cite{whitfield2023elicit}, as well as general search applications \cite{Venkatachary_2024}. These tools promise to transform scholarly settings by automating synthesis, accelerating systematic reviews from months to hours, performing intelligent citation analysis, and identifying research patterns and gaps \cite{10.1162/qss_a_00146,whitfield2023elicit}.
Moreover, emerging LLM-based research agents like DeepResearch \cite{deepresearch} and Google's AI Co-Scientist \cite{gottweis2025aicoscientist} are presented as able to autonomously conduct literature reviews and even generate research hypotheses.

The promise of automated research processes remains critically undermined by persistent limitations in the reliability of LLM-based systems.
While LLM-augmented scholarly tools generate fluent and authoritative-sounding outputs, they inherit the same ethical issues and harms as their base models \cite{10.1145/3442188.3445922,10.1145/3531146.3533088,kumar-etal-2023-language}. They particularly risk introducing errors due to their propensity to hallucinate information \cite{10.1145/3571730,10.1145/3703155,lin-etal-2022-truthfulqa,maynez-etal-2020-faithfulness}. These errors are difficult for scientists to detect \cite{bajpai-etal-2024-llms}, and persist even in Retrieval-Augmented Generation (RAG) systems, designed to be more reliable by grounding their generations in a corpus \cite{munikoti-etal-2024-evaluating,lewis2021retrievalaugmentedgenerationknowledgeintensivenlp}. These errors carry particularly high stakes in academic contexts where domain experts require precise information for literature reviews, peer assessment, and research design. Finally, LLMs can spread misinformation \cite{Kreps_McCain_Brundage_2022,10.1145/3544548.3581318}, spurring concerns that relying on ungrounded information systems will result in error propagation through scholarly discourse \cite{10.1145/3616863}.

Predicting erroneous outputs and understanding the full capabilities of LLMs is difficult \cite{10.1145/3531146.3533229}. Current mechanisms for trust and reliance calibration in LLM outputs are limited, and offer insufficient affordances for a setting like scholarly QA. For example, source citations for attribution in LLM responses can improve perceptions of trust, but have been found to sometimes be hallucinated or inaccurate in representing the source material~\cite{byun2024reference,narayanan2025search}. 
Similarly, uncertainty visualization methods like confidence highlighting can reduce over-reliance \cite{10.1145/3706598.3714097}, but rich explanatory information often increases cognitive load \cite{abdul2020cogam} without providing actionable paths for evidence validation .
Approaches combining sources with explanations are promising but remain vulnerable to an ``illusion of explanatory depth'' where users overestimate their understanding without actually verifying evidentiary support \cite{10.1145/3397481.3450644,10.1145/3706598.3714020}. However, fostering appropriate trust is not merely a matter of providing better information. Scholarly work occurs under time pressure and cognitive load---conditions that may prevent researchers from acting on their skepticism even when they recognize potential problems. Understanding how designs for trust and reliance calibration interact with these practical constraints is essential for developing effective scholarly AI tools.\looseness=-1

In this work, we design a novel provenance mechanism for LLM outputs in scholarly QA settings, grounded in the argumentation structures inherent to scholarly work. We build on research that shows how structured representations of claims and evidence can improve interpretability \cite{10.1145/3672608.3707811,10.1145/3580479}. Argumentation structures offer promise for scholarly QA because they mirror academic discourse structures, where claims are supported by evidence connected through warrants; scientific papers inherently follow these argumentation patterns \cite{lauscher-etal-2018-investigating,green-2014-towards}. Our system, \system, makes these implicit structures explicit through claim-evidence matching between a source document corpus and LLM responses to user queries for scholarly QA. It presents this provenance information via interface indicators that provide immediate visual feedback to the user by showing unsupported answer claims and omitted paper claims. Our design leverages domain experts' existing mental models of how scholarly arguments work, which has the potential to enable more efficient verification than generic explanation approaches. \looseness=-1

We evaluate \system and our argument-grounded source provenance approach through a within-subjects user study involving 26 domain experts recruited from a research organization. We compare our setup against a baseline of source citations for attribution, common in commercial LLM-based search tools. Our participants complete two scholarly tasks where they are asked to edit LLM-generated text after a QA session with the LLM in question. We measure three outcomes of provenance information presentation: subjective trust perception~\cite{10.3389/fcomp.2023.1096257}, self-confidence in participants' revised outputs, and behavioral reliance quantified via normalized Levenshtein edit distance between the original LLM-generated text and the participants' edited versions. 

Our results show that granular claim-evidence provenance information encourages more caution towards LLM outputs in scholarly settings. People trust in LLMs is significantly lower after using \system compared to baseline. However, this change does not translate to differences in perceived confidence or, importantly, changes in reliance behaviors. Experiential measures of usability and qualitative feedback suggest that, while helpful, this additional argument-grounding information clutters the interface and reduces usability, especially given the time constraints of a study setup. However, participants consistently appreciate the ethos of receiving this detailed breakdown of LLM arguments. We discuss the value of our argument-grounded source provenance approach for establishing trustworthiness of LLMs in this scholarly context, and discuss design implications for further reducing the cognitive load of presenting this additional information.

This work makes the following contributions to human-AI collaboration in information-intensive contexts:
\vspace{-.1cm}
\begin{itemize}
    \item The design and implementation of \system, a novel system that operationalizes argumentation structures for source provenance by decomposing and linking claims and evidence between LLM-generated answers and source documents.
    \item A flexible backend architecture for claim-evidence extraction that can serve as an interactive tool and as a framework for evaluating LLM trustworthiness in scholarly settings.\looseness=-1
    \item Empirical evidence from a within-subjects study showing the value of claim-evidence provenance in calibrating trust compared to standard source-citations from commercial LLMs.\looseness=-1
    \item Empirical evidence of a trust-behavior gap in scholarly LLM use, showing that reduced trust alone is insufficient to change reliance behaviors without addressing systemic constraints of time, usability, and cognitive resources.\looseness=-1
\end{itemize}

\section{Related Work}

\add{\subsection{Evidence-Based Text Generation, Attribution, and Provenance}}
\add{Recent surveys position LLM sourcing as a critically important design problem. \citet{schreieder2025attributioncitationquotationsurvey} survey evidence-based text generation across 134 systems that use source material to ground model outputs in external evidence. \citet{pang2025largelanguagemodelsourcing} distinguish provenance at the levels of model authorship, model structure, training data, and external data, and separate prior-based approaches that embed explicit markers from posterior-based approaches that infer provenance from observed behavior. Within this broader sourcing landscape, \citet{li2023surveylargelanguagemodels} focus on external-data sourcing for question answering, formalizing attribution as mapping each answer statement to one or more cited passages and evaluating methods in terms of citation coverage (recall) and sufficiency (precision).\looseness=-1

Across these surveys, most systems attach provenance at document, paragraph, or sentence granularity and treat attribution primarily as a backend or benchmarking problem: how to retrieve better evidence, assign more accurate citations, or define more faithful automatic metrics. Coarse-grained citations lead to familiar failure modes such as granularity errors, mistaken synthesis across sources, and hallucinated statements when complex answers are supported only by flat sentence-level links \cite{li2023surveylargelanguagemodels}, and little work has studied how experts actually use provenance cues in context. In \citet{pang2025largelanguagemodelsourcing}'s terms, our system is a posterior external-data sourcing system: we do not modify model weights or embed watermarks, but instead infer and expose how an LLM answer relates to a fixed corpus of scholarly articles at interaction time. Building on the attribution formulation of statement-plus-citations \cite{li2023surveylargelanguagemodels}, we instantiate provenance at a finer granularity by decomposing both papers and answers into discrete claims, aligning answer claims to paper claims and evidence snippets, and making omissions and mismatches explicit.}\looseness=-1

\subsection{LLMs and Explanations}
Prior work in human-centered AI asks not only what information LLMs should expose, but also how such cues can shape people's trust perceptions and reliance behaviors. Approaches towards transparency, including model reporting, evaluations, explanations, and communicated uncertainty, can be a means to support people in appropriate trust calibration, particularly when tailored to stakeholder goals and contexts \cite{liao2024ai}. Recent behavioral studies probe which cues actually foster appropriate reliance. Uncertainty cues were found to help reduce overreliance \cite{10.1145/3630106.3658941} and confidence highlighting were found to help users catch errors \cite{10.1145/3706598.3714082} in two recent lab studies. Earlier work established that confidence scores and explanations interact in non-trivial ways, with explanations improving trust calibration primarily when combined with appropriate confidence information~\cite{10.1145/3351095.3372852}.\looseness=-1

It is important to consider which kinds of transparency impact behavior. \citet{10.1145/3706598.3714020} found that explanations overall tend to increase reliance on both correct and incorrect answers, while verifiable sources help reduce overreliance when the model is wrong and support appropriate reliance when it is right; they also identify ``inconsistencies'' as a distinct unreliability cue in LLM outputs \cite{10.1145/3706598.3714020}. In controlled reliance interventions, simple, persistent disclaimers can be effective, while token-level uncertainty cues or removing direct answers reduce overreliance but often at time costs and without reliably improving appropriate reliance \cite{10.1145/3706598.3714097}. For domain experts using retrieval-augmented generation (RAG) systems, surfacing sources and uncertainty interacts with users' verification practices and trust, which shows the importance of provenance cues in expert workflows \cite{10.1145/3706599.3719985}; such interventions can also assist domain experts in identifying confabulations in RAG-based systems \cite{10.1145/3706599.3720249}. 

\subsection{Argument Structures in Textual Contexts}
Our interface foregrounds claims and the evidence that supports them based on argumentation theory. In predictive-advice settings, structuring justifications with \citet{Toulmin_1958}'s argument structure components (data, warrants, backings, and rebuttals) selectively strengthens distinct trusting beliefs, suggesting that showing what the claim is and why it might not hold can calibrate trust more effectively than undifferentiated explanations \cite{10.1145/3580479}. In scholarly writing specifically, argument structure underlies rhetorical function. Recent work augmented a scientific corpus with argumentative components and find that coupling argument extraction with rhetorical tasks in multi-task machine learning improves performance, and that argument components are most tightly linked to discourse roles \cite{lauscher-etal-2018-investigating}. A survey of argument mining for scholarly document processing further establishes that scientific texts contain rich argumentative structure amenable to computational extraction, while identifying open challenges in connecting mined arguments to downstream user tasks~\cite{al-khatib-etal-2021-argument}.We draw on this finding to propose that claims and evidence are the right unit to expose for scholarly sense-making. 

We also build on work that evaluates argument-based explanations as user-facing artifacts. In medical QA, a recent study mined argument components and assessed explanation structure via graph patterns (e.g., missing premises, inconsistent support/attack) \cite{10.1145/3672608.3707811}. They found that people benefit when explanations are explicitly organized as arguments rather than free-form text. Beyond professional domains, an educational psychology study shows that recomposing arguments with Toulmin elements can measurably improve critical-thinking skills \cite{10.1145/3578837.3578869}.

\subsection{Scholarly Question Answering}
Work on scholarly question answering (QA) clarifies both the task demands of answering research questions from papers and the interface signals needed for credible use. Scholarly corpora such as PubMedQA \cite{jin-etal-2019-pubmedqa} and Qasper \cite{dasigi-etal-2021-dataset} establish that answering researcher-style questions requires reasoning over long, technical texts rather than factoids. More recent studies broaden the space to knowledge-graph QA \cite{Auer2023}, large-scale science QA \cite{Saikh2022}, expert-authored long-form questions with attributed answers \cite{malaviya-etal-2024-expertqa}, multi-document / multimodal settings \cite{li-etal-2024-m3sciqa,pramanick2025spiqadatasetmultimodalquestion}, and exam-style free-response evaluation \cite{dinh-etal-2024-sciex}. Scholarly QA evaluation show limitations of LLM-based QA systems. Long-context models still degrade with text distance \cite{hilgert-etal-2024-evaluating}, retrieval-augmented models can fabricate supporting evidence in science tasks \cite{munikoti-etal-2024-evaluating}, and domain experts judge model outputs as coherent yet inconsistently accurate \cite{peskoff-stewart-2023-credible}. \citet{10.1145/3772318.3791843} developed an expert-derived schema identifying specific error types in scholarly QA, which goes beyond inaccuracies and hallucinations to describe issues with synthesis, formatting, question interpretation, and completeness. These performance issues motivate attribution-first interfaces that make evidence not only available to users, but also intelligible.

Argument structures are important to several works on Scholarly QA. Scientific claim verification shows the importance of aligning claims with cited evidence in open domains \cite{wadden-etal-2022-scifact}. More recently, SciClaimHunt introduces large-scale scientific claim-verification resources \cite{kumar2025sciclaimhuntlargedatasetevidencebased}. Such argumentation-forward resources show that scholarly discourse is naturally structured around claims, supports, and rebuttals \cite{ruggeri-etal-2023-dataset}, and that LLMs evaluated as science communicators can appear persuasive while remaining unreliable \cite{bajpai-etal-2024-llms}, which shows the importance of foregrounding verifiable sources. Finally, HCI perspectives urge centering domain experts' values and workflows in NLP tools \cite{10.1145/3532106.3533483,10.1007/s10115-024-02212-5}, and QASA \cite{10.5555/3618408.3619195} contributes a scholarly question taxonomy and full-stack reasoning setup that our work leverages. Our research extends this line of inquiry by looking at how a claim-evidence-based user interface affects trust perceptions and behavioral reliance during scholarly writing and critique.\looseness=-1

\section{\system: a Scholarly Source Provenance System}
We implemented a novel system to investigate how argument-grounded provenance affects user trust and reliance in LLM-based scholarly QA. The system generates and displays claim-evidence structures from both a corpus of source documents and real-time LLM-generated answers. Below, we first describe our backend architecture, in particular our novel argument extraction and matching engine for determining source provenance; followed by our design goals and frontend interface.

\subsection{Backend: Argument Extraction and Matching for Grounding Provenance} \label{sec:claim-evidence_extraction}
To generate the argument-grounded provenance annotations intended to help users appropriately calibrate their trust and reliance, we developed an Argument Extraction Engine that combines three approaches to claim and evidence extraction, each offering different tradeoffs between computational cost and semantic capability. \HighCost{LLM-based extraction} leverages LLMs' semantic understanding to identify claims and evidence based on few-shot prompting, with their linguistic understanding offering a more nuanced interpretation of scientific discourse. While computationally expensive, this approach produces natural language representations that align with human categorization and can capture semantic meaning beyond surface-level text similarity \cite{pham-etal-2024-topicgpt}. \LowCost{Similarity-based extraction} uses a sentence-transformer model and cosine similarity for rapid filtering and deduplication. This lightweight approach offers speed, interpretability, and reliability for high-volume operations where semantic nuance is less critical. \MediumCost{Retrieval-Augmented Generation (RAG)} combines retrieval efficiency with LLM semantic understanding by first filtering content using similarity search, then applying LLM processing to the reduced set. \add{This hybrid approach reduces computational cost compared to applying LLMs to full documents by limiting the candidate set}. 

\begin{figure*}[htpb!]
    \centering
    \includegraphics[width=\linewidth]{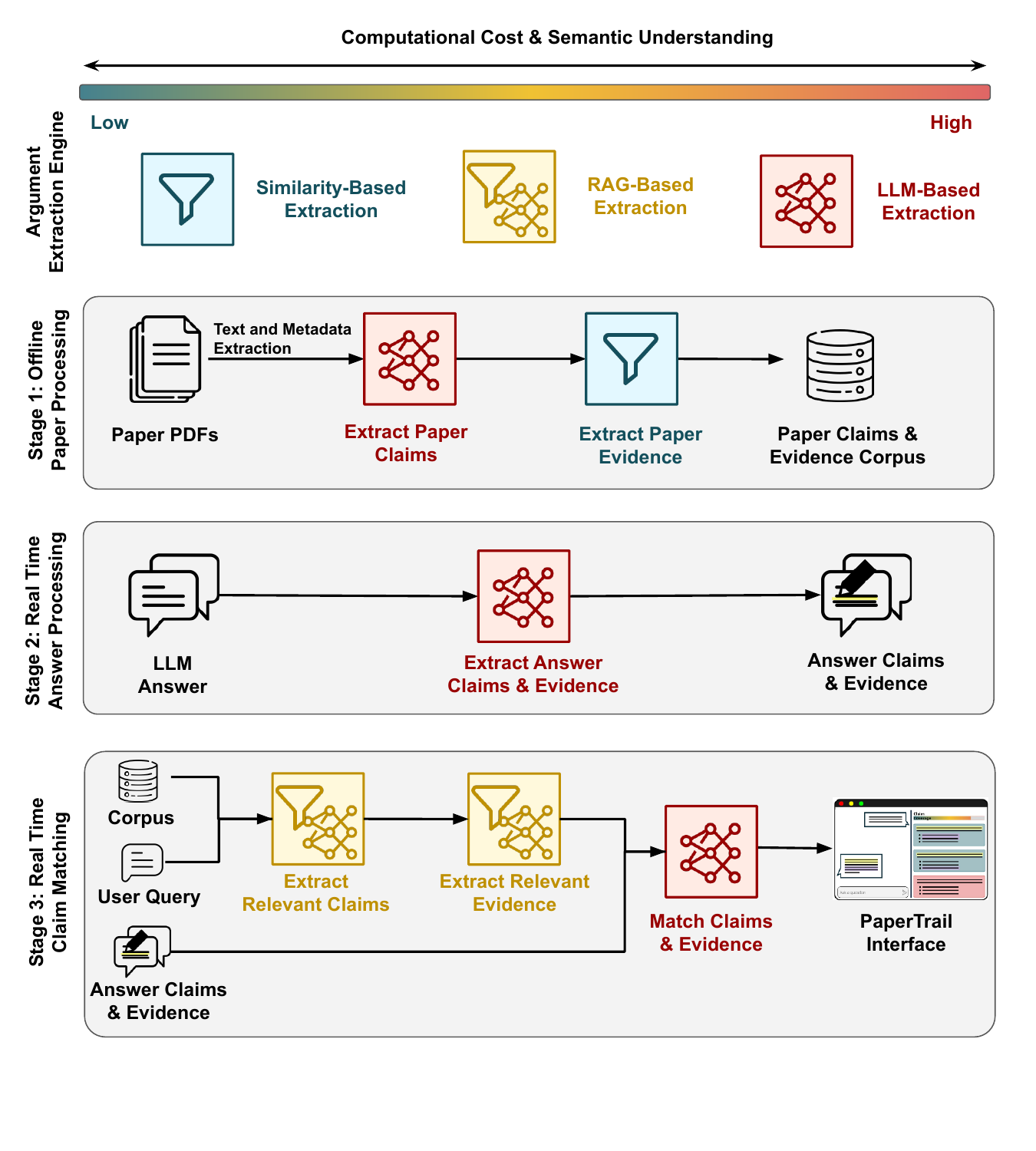}
     \vspace{-5pc}
     \Description{A detailed diagram of the Argument Extraction Engine's three-stage pipeline, organized as a matrix. The top row shows the three extraction methods arranged along a horizontal axis labeled ``Computational Cost \& Semantic Understanding'' from low to high: similarity-based extraction (teal), RAG-based extraction (gold), and LLM-based extraction (red). Below, three rows represent the pipeline stages. Stage 1 (Offline Paper Processing) shows paper PDFs undergoing text and metadata extraction, then LLM-based claim extraction (red), followed by similarity-based evidence extraction (teal), producing a paper claims and evidence corpus. Stage 2 (Real-Time Answer Processing) shows an LLM answer undergoing LLM-based claim and evidence extraction (red), producing answer claims and evidence. Stage 3 (Real-Time Claim Matching) takes the corpus, user query, and answer claims as inputs; applies RAG-based extraction (gold) to extract relevant claims and evidence; then matches claims and evidence to produce output for the PaperTrail interface. Color coding throughout follows the computational cost gradient: teal for lowest cost, gold for medium, and red for highest.}
    \caption{The Argument Extraction Engine (top) provides \add{three extraction methods with different computational cost and accuracy tradeoffs.} Colors follow a computational cost gradient: \HighCost{red} indicates LLM-based extraction (highest computational cost), \MediumCost{gold} represents RAG-based extraction (medium cost), and \LowCost{teal} denotes similarity-based extraction (lowest cost). \add{We deploy these methods strategically across pipeline stages based on design-time considerations}: (1) offline paper-level information extraction that preprocesses research papers into structured claims and evidence; (2) real-time answer-level extraction that decomposes LLM-generated answers into claims and supporting evidence; and (3) real-time claim-evidence matching that uses retrieval-augmented generation (RAG) to filter and align relevant paper claims with answer claims, producing source provenance indicators.}
    \label{fig:backend}
\end{figure*}

We deploy these methods \add{at different pipeline stages based on stage-specific requirements: whether a user query is available to guide relevance filtering, whether processing occurs offline or in real-time, and the volume of text to be processed. The resulting} three-stage pipeline consists of: offline paper-level extraction of claims and evidence for a source document corpus (Section \ref{sec:paper-level_claim-evidence}); real-time answer-level extraction for LLM-generated scholarly QA answers (Section \ref{sec:answer-level_claim-evidence}); and real-time claim-evidence matching to calculate argument-grounded source provenance between scholarly documents and real-time LLM answers in QA (Section \ref{sec:claim-evidence_matching}). The pipeline is served by a Flask web server that provides a RESTful API, session management, interaction logging, and static content delivery. See Figure \ref{fig:backend} for an illustration of the backend stages.

\subsubsection{Stage 1: Paper-Level Claim-Evidence Extraction}\label{sec:paper-level_claim-evidence}
We construct a corpus of scholarly documents structured into claims and evidence through one-time offline preprocessing, where computational cost is less constrained than in real-time operations. We preprocess each document by extracting the text using PyMuPDF v1.23.5,\footnote{\url{https://pypi.org/project/PyMuPDF/}, GNU Affero General Public License} followed by manual validation to ensure accuracy. \add{We reviewed the preprocessed text against the original PDFs to check for the correctness of extracted mathematical notation; correct preservation of paragraph boundaries; and accurate handling of hyphenated words across line breaks.}
The plain text is programmatically segmented into sections and paragraphs to maintain contextual coherence and to allow for detailed claim extraction.

For claim extraction in this offline setting, we use \HighCost{LLM-based extraction} where each paragraph is provided as context to Gemini 2.5 Pro \cite{comanici2025gemini25pushingfrontier} using a few-shot prompt inspired by the work of \citet{kumar2025sciclaimhuntlargedatasetevidencebased} and \citet{Toulmin_1958}'s argumentation model. This prompt provides 10 examples of claims randomly selected from the SciClaimHunt dataset~\cite{kumar2025sciclaimhuntlargedatasetevidencebased} and instructs the model to identify and extract distinct, verifiable scientific claims. \add{Following \citet{kumar2025sciclaimhuntlargedatasetevidencebased}, we specify that claims should be \textit{atomic}, \textit{faithful}, and \textit{decontextualized}. Drawing on \citet{Toulmin_1958}'s argumentation framework, we additionally require claims to be \textit{verifiable} (checkable against evidence) and \textit{declarative} (statements rather than questions or method descriptions). \citet{kumar2025sciclaimhuntlargedatasetevidencebased} validated these criteria through human annotation of 100 claims, achieving inter-annotator agreement of $\alpha = 0.69$--$0.78$ across dimensions. We adopted their methodology and qualitatively spot-checked claims extracted from one paragraph from each of the Introduction, Methods, and Results sections from each paper. 
These sections represent distinct discourse functions, where Introductions contain motivational and background claims, Methods contain procedural claims, and Results contain empirical findings; we wanted to verify that the extraction prompt handled this variation appropriately. This spot-check confirmed that extracted claims generally satisfied the five criteria from \cite{kumar2025sciclaimhuntlargedatasetevidencebased,Toulmin_1958} noted above. We cover details on our quantitative evaluation of this approach in Section \ref{sec:offline-eval} below.}

We use LLM-based claim extraction here despite its computational cost for two reasons. First, without a query to inform relevance filtering, we must extract all potentially relevant claims from the corpus comprehensively. Similarity-based filtering or RAG approaches require a query as a reference point for relevance scoring. Second, the offline nature of this preprocessing step makes the computational expense acceptable, as it occurs once per document rather than in real time. While we could have eliminated the preprocessing step and only extracted paper claims in real-time using the user's query to identify relevant paper claims, this would introduce additional latency into every user interaction. Pre-extracting all claims allows us to more quickly identify relevant claims by searching an already-processed corpus. This shifts the computational burden from synchronous user-facing operations to asynchronous preprocessing. It also makes the interactions more consistent across users, because offline preprocessing creates a single ground truth. Additionally, pre-extraction enables reproducible results and allows the claim-evidence corpus to be versioned and audited independently of the real-time query processing pipeline.

Finally, in the evidence retrieval stage, we use \LowCost{Similarity-based extraction}, where the claims extracted previously are used as queries to find their supporting evidence within the source text. Sentences exceeding a similarity threshold of 0.75 (based on guidance from \citet{kumar2025sciclaimhuntlargedatasetevidencebased}) are considered candidate evidence for a given claim. To improve the readability of the extracted evidence, the preceding and subsequent sentences surrounding each evidence snippet are included as context to reconstruct coherent snippets. We use low-cost similarity-based extraction for evidence retrieval because this stage serves only as an initial filtering step to identify potentially relevant passages. The semantic relevance of this evidence to a user's specific information needs is later determined by LLM-based processing during the real-time claim-evidence matching phase (Section \ref{sec:claim-evidence_matching}), where the user's query provides the necessary context for selecting the most pertinent evidence. At this preprocessing stage, we simply need to establish which text segments have any topical relationship to each claim—capturing passages that mention the same concepts, entities, or phenomena. This broad initial retrieval ensures comprehensive coverage while deferring the computationally expensive task of determining contextual relevance until it can be informed by the user's actual query.\looseness=-1

The output of this preprocessing pipeline is a single JSON file containing a structured list of all paper claims including each claim's associated evidence, citation, and section name. This file is loaded by the backend server at startup and serves as the ground-truth knowledge base for \system. 

\subsubsection{Stage 2: Answer-Level Claim-Evidence Extraction}\label{sec:answer-level_claim-evidence}
Real-time answer-level extraction involves two distinct models serving different roles: an \textbf{answerer LLM} that responds to user questions, and an \HighCost{extraction LLM} that decomposes these answers into claims and evidence. This separation ensures that the answer generation remains focused on content quality while the extraction process maintains structural consistency with how paper claims were processed.\looseness=-1

When a user asks a question during the interactive session, the \textbf{answerer LLM} first generates a complete answer based on the query, conversational history, and task context. \add{In our study configuration, the answerer LLM receives the source documents as context alongside the query, similar to document-grounded question answering in commercial LLM interfaces where users upload PDFs and ask questions about their content. While this differs from RAG-based systems that retrieve relevant passages based on the query, \system's claim-evidence extraction and matching stages are model-agnostic and would function identically with RAG-generated answers.} This model operates without additional structural constraints, which aligns the outputs to familiar commercial LLM chat experiences: mimicking the natural question-answering flow users encounter when uploading PDFs to commercial models and asking questions about their content. This separation between answer generation and claim extraction also makes our system model-agnostic; it is capable of analyzing outputs from any LLM rather than requiring a specific architecture or training approach. \add{In standard RAG deployments, \system would function identically: the claim-evidence extraction and matching stages operate on the generated answer regardless of how that answer was produced. The key difference from typical RAG interfaces is granularity—while RAG systems often display retrieved passages as coarse-grained attribution, \system decomposes both the answer and source documents into discrete claims, which allows for verification of specific assertions rather than entire passages.}

Once the answer is generated, the secondary \HighCost{extraction LLM} performs the claim-evidence decomposition using the same definition of claims as the prompts used for paper-level extraction. In addition to the prompt, which is modified to also request supporting evidence to be identified for each claim, the LLM receives the complete answer text along with a JSON schema that enforces structured output (Google's Gemini API allows for a schema specification to be passed as an argument that dictates how the output should be structured, which guarantees usable information extraction). Since answers are much shorter than full paper texts, claim and evidence extraction occurs in a single pass without a separate evidence retrieval step. We use \HighCost{LLM-based extraction} for this stage despite its computational cost because the context is significantly shorter than full papers, making the processing time acceptable for real-time interaction. Additionally, we can extract both claims and evidence in a single LLM call rather than requiring separate passes. \looseness=-1

Following extraction, the system performs span annotation using the NLTK punkt tokenizer to segment the text into sentences. Then, a programmatic matching function locates the precise character positions of each claim and evidence piece identified by the extraction LLM, which maps the structured output back to specific text spans in the original answer. This lightweight post-processing step enables precise text highlighting required for the interactive user interface, where users need to see exactly which portions of the answer correspond to specific claims.

\subsubsection{Stage 3: Claim-Evidence Matching}\label{sec:claim-evidence_matching}
Our system matches the claims and evidence across the corpus of source papers and an individual LLM answer in QA using the structured claim-evidence representations extracted from the previous two steps. The goal of this stage is to improve the trustworthiness of the system by creating both global explanations (overall claim coverage) and local explanations (individual claim provenance) for the scholarly setting. 

The matching process uses \MediumCost{RAG-based extraction} to identify relevant claims and evidence from the paper. In this RAG pipeline, the corpus consists of all paper-level claims and their supporting evidence extracted offline from source documents. The query is the user's asked question. Similarity-based retrieval first filters the complete paper claim corpus to identify candidates relevant to the user's question (using cosine similarity between the SPECTER embeddings of the question and each paper claim). The LLM is then prompted to select from the list of extracted claims the most relevant ones to the query. The same \MediumCost{RAG approach} is applied to evidence selection: for each relevant claim, its supporting evidence forms a sub-corpus that is searched using the query, with similarity-based retrieval filtering candidates and the LLM performing final selection of the most relevant evidence passages. We use \MediumCost{RAG-based extraction} in this step because the LLM provides semantic capabilities to disambiguate between superficially similar but conceptually different claims and evidence, while the initial similarity-based retrieval dramatically reduces the search space from hundreds of claims/evidence to a manageable set of candidates.

Next, the LLM is prompted to perform claim-to-claim matching. The model is tasked with comparing the list of answer claims to the filtered list of relevant source papers' claims, and to find the most semantically equivalent pairs. This step benefits from the model's more nuanced language understanding to map the answerer LLM's generated assertions in scholarly QA to their ground-truth counterparts in the source literature, and avoids spurious connections that can occur when using simple cosine-similarity matching.\looseness=-1

Finally, the evidence from the answer is verified using cosine similarity. To help users calibrate their trust and reliance on the system's output, we set a permissive cosine similarity threshold of $<0.55$ to flag \add{potentially} unsupported evidence. This threshold value was selected \add{through iterative testing on five held-out examples} to satisfy two requirements: first, to identify answers that diverge substantially from the source material; and second, to avoid alert fatigue and acknowledging that relevant evidence in an answer may not be present in the filtered set of paper evidence. Our goal is to help users form accurate mental models of the system's capabilities \cite{10.1145/3290605.3300233} to support appropriate reliance and mitigate both overtrust and undertrust of the answerer LLM \cite{doi:10.1518/hfes.46.1.5030392}, and to manage user expectations of imperfection \cite{10.1145/3290605.3300641}.

\subsubsection{\add{Implementation Details}}\label{sec:implementation}
\add{Our system uses Gemini 2.5 Pro (gemini-2.5-pro-preview-05-06) for all LLM-based extraction and matching operations with temperature set to 1.0. We enforced structured JSON output using Gemini's response\_json\_schema parameter, which guarantees responses conform to our claim-evidence schema. We used the sentence-transformer model SPECTER~\cite{specter2020cohan} for all similarity-based operations, which we selected for its optimization on scientific text. Average end-to-end response latency was around 90 seconds per query, primarily due to sequential LLM calls in Stages 2 and 3. Full prompts are provided in Appendix~\ref{app:prompts}.}

\subsection{Frontend Interface} \label{sec:interface}
\begin{figure*}[htpb!]
    \centering
    \includegraphics[width=0.98\linewidth]{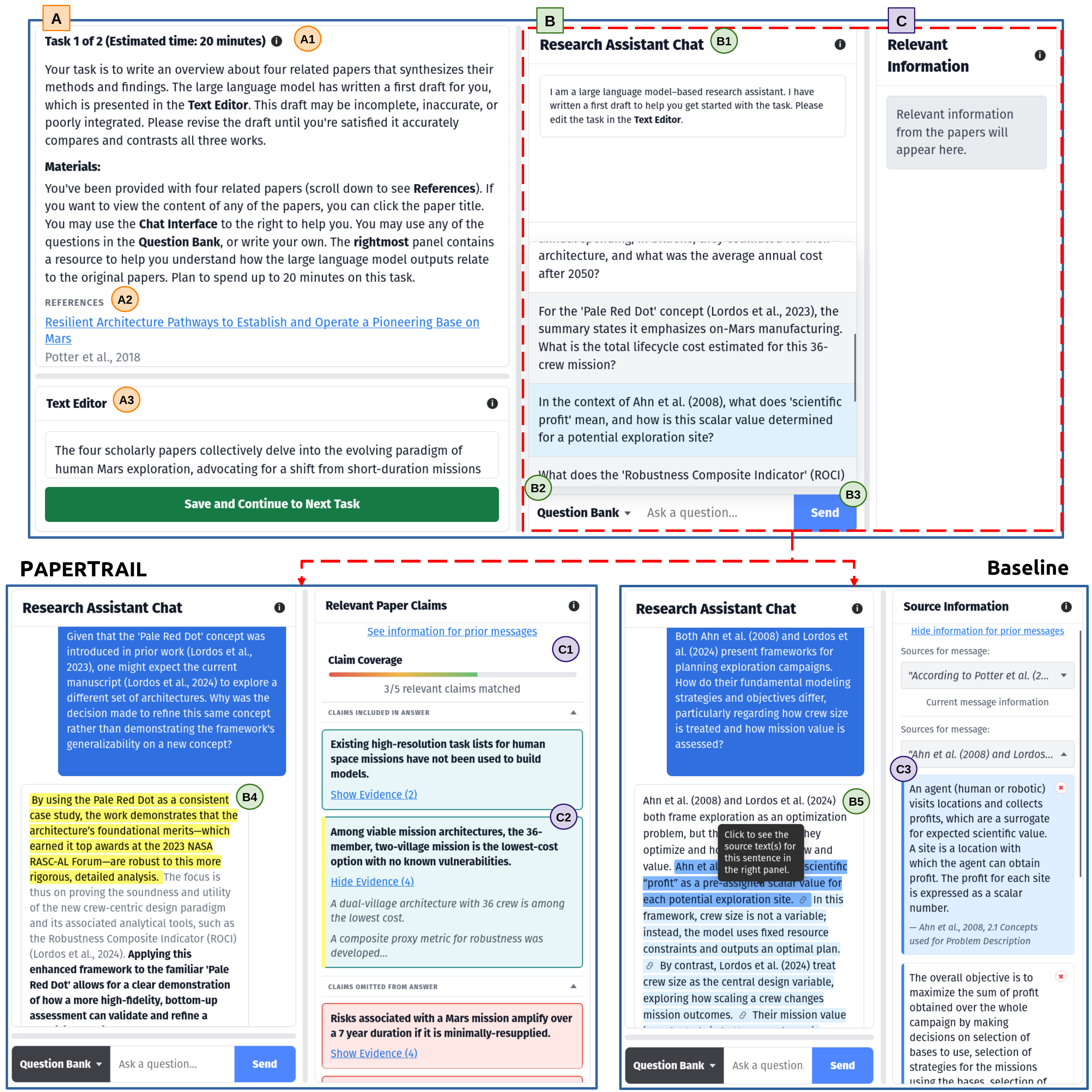}
    \Description{A composite screenshot of the PaperTrail user interface in three parts. The top portion shows the shared three-panel layout: the Left Panel (A) contains the task workspace with task context instructions (A1), a references list of four Mars exploration papers (A2), and a text editor with an LLM-generated draft (A3); the Middle Panel (B) shows the chat interface (B1) with a conversation history, a question bank dropdown (B2), and a send button (B3); the Right Panel (C) is a placeholder for provenance information. The bottom portion shows side-by-side comparisons of the two study conditions. On the left, the PaperTrail condition displays answer text with highlighted interactive answer claims (B4) in the middle panel, while the right panel shows a ``Claim Coverage'' bar indicating 3 of 5 relevant claims matched (C1), followed by paper claim cards (C2) organized into ``Claims included in answer'' (teal cards) and ``Claims omitted from answer'' (red cards), each with expandable evidence sections. A selected claim is highlighted with a yellow vertical bar. On the right, the baseline condition shows sentence-level source highlights (B5) in the middle panel with a tooltip prompting users to click to see sources; the right panel displays verbatim paper source text (C3) organized by citation under expandable dropdown sections.}
    \caption{The user interface comprises three main panels. The general layout, shown at the top, consists of the \LeftPanel, the \MiddlePanel, and the \RightPanel. The \LeftPanel contains the user's main workspace, including the \TaskContext, a \ReferencesList, and the \TextEditor. The \MiddlePanel serves as the \ChatInterface, which includes a \QuestionBank and \ChatControls. The \RightPanel displays information provenance, with its content changing based on the condition. The lower half shows the differences between the conditions. In the \system interface (left), the \MiddlePanel shows interactive \AnswerClaims. When a user clicks a claim, the corresponding \PaperClaim is highlighted in the \RightPanel, which also shows the overall \ClaimCoverage for the LLM's answer. In the baseline interface (right), the \MiddlePanel contains a sentence-level \SourceHighlight. Clicking this highlight surfaces the verbatim \PaperSource text in the \RightPanel.\looseness-1}
    \label{fig:system-diagram}
\end{figure*}

Our user interface is a three-panel web application designed to support scholarly question answering (QA) tasks (Figure \ref{fig:system-diagram}). The \LeftPanel contains the user's primary task workspace, the \MiddlePanel contains the LLM-based scholarly QA chat, and the \RightPanel provides provenance information based on claim-evidence matching. This section presents our four design goals and then describes how these goals are instantiated in the interface components.

\subsubsection{Design Goals}
Our interface for \system is based on four design goals informed by challenges of source provenance in scholarly settings. Overall, these goals address tension between providing comprehensive provenance information while maintaining usability for domain experts, who are engaged in complex analytical tasks.\looseness=-1 

\textbf{DG1: Support graduated cognitive engagement.} Our goal is to structure access to provenance such that users can get a collective sense of the main provenance metric---number of claim-evidence matches---and dive into details as needed. This pattern builds on \citet{10.5555/832277.834354}'s visual information-seeking mantra: ``overview first, zoom and filter, then details-on-demand''. This graduated cognitive engagement is helpful for scholarly tasks, which often involve what \citet{10.1145/1121949.1121979} calls ``exploratory search'': the scholar moves beyond fact retrieval to support deeper learning and investigation. In their foundational work on a cognitive task analysis of literature search, \citet{pirollicard2005} outline several cognitive processes under foraging and sensemaking loops that help people gradually synthesize information. This design goal for our intended interface reflects their model by connecting their ``shoebox'' stage (the collected source papers) and ``evidence file'' stage (the extracted claims and evidence) \cite{5333878,pirollicard2005}. Our interface allows users to operate at different levels of the sensemaking process, from a high-level answer (overview) to claim-level annotations (zoom/filter) and finally to the source text itself (details-on-demand). 

\textbf{DG2. Minimize interaction overhead while preserving exploration depth.} 
Source provenance inherently adds more elements to an already interaction-heavy interface, e.g., by providing more links to click and more information modalities to handle. To keep provenance manageable, we follow design principles for Coordinated Multiple Views (CMV) \cite{10.1145/345513.345271,4269947}. Specifically, we adhere to \citet{10.1145/345513.345271}'s guidelines by optimizing for space and time resources with a multi-pane layout, ensuring self-evidence by using brushing-and-linking to make relationships between claims and sources clear, and supporting attention management with selective information hiding. Following \citet{pirollicard2005}'s information foraging cost structures, our aim is to provide comprehensive verification capabilities while reducing engagement burden. In our interface, automatic claim highlighting, structured claim-evidence decomposition, and focused interaction modes are designed to reduce the costs of scanning, recognizing, and selecting information, respectively.

\textbf{DG3. Enable flexible verification workflows.}
Experts approach the same data using a variety of strategies \cite{5333878}, and opportunistically combine bottom-up and top-down processes based on emerging insights and verification needs. For example, an expert might ``find a clue, and follow the trail'' \cite{5333878}, a non-linear process that requires flexibility to support. This also aligns with principles of exploratory search, which emphasize iteration and discovery over linear lookup \cite{10.1145/1121949.1121979}. Our interface avoids enforcing a single verification pathway. Users can navigate bidirectionally between claims and evidence, which supports both top-down hypothesis checking (starting from answer claims to find supporting evidence) and bottom-up evidence discovery (exploring paper claims to identify gaps in the answer).

\textbf{DG4. Align with scholarly mental models and human-AI interaction guidance.}
To be effective, an intelligent system must align with its users' existing knowledge structures and meet their expectations for interaction. Experts 
develop mental models of their domain that rely on structural and causal features rather than superficial ones \cite{10.5555/166431.166438}. Our system reflects inherent argumentative structures of scientific discourse \cite{ruggeri-etal-2023-dataset,wadden-etal-2022-scifact,kumar2025sciclaimhuntlargedatasetevidencebased,lauscher-etal-2018-investigating} by decomposing information into claims and evidence, so that it can leverage researchers' existing mental models.

\subsubsection{Interface Layout and Features}
The user interface is a single-page web application built with React and Redux Toolkit for state management. It presents a three-panel, resizable layout so users can keep multiple sources of information in view without context switching, a design decision aligned with \textbf{DG1} and \textbf{DG2} by maintaining continuity of attention across related elements \cite{10.1145/345513.345271,4269947}. The panels provide the task context (\LeftPanel), the chat interface (\MiddlePanel), and the source provenance (\RightPanel). The panels are synchronized through a lightweight event bus, which allows brushing, linking, and coordinated navigation \cite{4269947,4389006}. Panel boundaries are resizable (horizontal and vertical splits where present), supporting DG3's flexible verification workflows. Components use nested scrolling so panes and in-panel sections can be scrolled independently, enabling DG1's graduated cognitive engagement. User selections auto-scroll linked panels to the relevant context, implementing DG2's principle of minimizing interaction overhead through coordinated views. Pervasive tooltips define sections and describe functionality, supporting DG1's details-on-demand pattern.

The \LeftPanel contains the main task environment, which includes the \TaskContext and \TextEditor. The \TaskContext provides an overview of the task instructions and an explanation of the resources available to the user, including a \ReferencesList of source documents relevant for the task. This list of references is clickable and opens up a PDF viewer which superimposes the clicked reference PDF over the entire window. The left panel maintains proximity between the task description and the user's own writing space, the \TextEditor, which reduces the cognitive burden of switching between context and text editing (\textbf{DG1}). The \TextEditor contains a placeholder response to the scholarly task at hand generated by an LLM. Users can edit this text as they see fit, especially once they have used the scholarly QA interface in the \MiddlePanel. The components within the panel are vertically resizable so that the user can proportionally enlarge the task description or the editor when needed, which supports flexibility (\textbf{DG3}): some users may rely heavily on the task prompt while writing, while others may minimize it entirely after an initial read.\looseness=-1

The \MiddlePanel comprises the \ChatInterface where users can ask questions about the source documents used for the task (i.e., our scholarly QA setting), and use this information to edit the text in the \TextEditor. We provide a \QuestionBank, a set of pre-written questions relevant for the scholarly context that the user can simply click and select to request a response (\textbf{DG1,DG2}). We also include \ChatControls, a Send/Stop control that becomes \emph{Stop} while the system is waiting for the LLM response. This affords user control in deciding when to wait for an annotated LLM response with \AnswerClaims or switch strategies or queries (\textbf{DG3}). Users can read a generated response at a glance, or inspect more deeply using embedded annotations (\textbf{DG1,DG3}). Clicking on an answer claim engages the \RightPanel. All other text in the answer except for supporting evidence is grayed out to help the user focus on the selected claim (\textbf{DG2}). \looseness=-1

The \RightPanel 
presents \PaperClaims, which are claims from the paper that are relevant to the question asked; the supporting evidence for each claim is also provided (\textbf{DG4}). Paper claims which match with claims made in the answer are under ``Claims included in answer'' and are presented in teal cards. Paper claims that do not have a match in the answer are under ``Claims omitted from answer'' and are presented in red cards. Claims whose match has been selected in the \MiddlePanel are highlighted using a vertical yellow bar (\textbf{DG2,DG3}). The right panel also contains a global overview \ClaimCoverage, a horizontal bar indicator that represents the number of relevant claims from the source documents included in the LLM's QA response (\textbf{DG1}). The indicator is colored from red to teal, and these colors correspond claim inclusion/exclusion colors of each claim-evidence provenance card. 
Each provenance card is linked to a specific message from the chatbot, creating a labeled, collapsible section for each turn so information from prior messages persists and can be revisited rather than disappearing (\textbf{DG2,DG3)}. Within each section, cards can be expanded or removed to manage clutter. The interface applies selective information hiding and coordinated-view synchronization (brushing/linking with auto-scroll) \cite{10.1145/345513.345271,10.5555/832277.834354,4269947,4389006}.

\section{\add{Offline Evaluation for Backend Approach}}\label{sec:offline-eval}
\add{We conducted an offline evaluation to assess the quality of our claim extraction approach. 
No gold-standard benchmark exists for evaluating LLM-based claim extraction that generates a complete set of atomic claims (rather than extracting verbatim text spans) from scientific texts---existing datasets target either non-exhaustive sentence-level claim classification~\cite{achakulvisut2020claimextractionbiomedicalpublications,magnusson-friedman-2021-extracting} or claim verification~\cite{kumar2025sciclaimhuntlargedatasetevidencebased,alvarez-etal-2024-zero,wadden-etal-2022-scifact}. Therefore, we adapted two related datasets to approximate an extraction evaluation.}\looseness=-1

\add{\subsection{Evaluation Datasets}}
\add{
\subsubsection{SciClaimHunt}
\citet{kumar2025sciclaimhuntlargedatasetevidencebased} constructed a dataset of scientific claims generated from research paper paragraphs using few-shot prompting with Llama-2-13B, focusing on claims that could be verified against textual evidence rather than exhaustive extraction. Since the dataset provides paper-level but not paragraph-level claim mappings, we matched each claim to its source paragraph by first searching for exact string matches against sentences, then using SPECTER embeddings \cite{specter2020cohan} to find the most similar sentence for remaining claims. We sampled 50 paragraphs with at least 3 associated claims, filtering for samples with sufficient annotation density to enable meaningful comparison with our exhaustive extraction approach. We used an additional 50-sample holdout set from SciClaimHunt while developing our procedure to avoid overfitting.

\subsubsection{BioClaimDetect}
\citet{achakulvisut2020claimextractionbiomedicalpublications} released a human-annotated dataset of abstracts in the biomedical domain with sentence-level claim annotations. Annotators labeled entire sentences as claims without decomposing them into atomic units. We randomly sampled 50 abstracts from their test set,
and used the full abstract text as input and the annotated claim sentences as reference claims.\looseness=-1
}

\add{\subsection{Evaluation Procedure}}
\add{
For each sample, we applied our claim extraction prompt using Gemini-2.5-Pro with 10-shot examples drawn from their respective datasets. We embedded both reference and extracted claims using SPECTER and computed pairwise cosine similarities. A reference claim was considered ``matched'' if any extracted claim exceeded a similarity threshold of $\tau=0.9$; likewise for extracted claims matching references. This conservative threshold ensures matched claims are semantically near-equivalent rather than only topically related.\looseness=-1

We define recall as the proportion of reference claims matched by at least one extracted claim, measuring coverage of benchmark claims; precision as the proportion of extracted claims matched by at least one reference claim, measuring extraction accuracy; and F1 (the harmonic mean of precision and recall). Note that neither dataset used for this evaluation provides exhaustive atomic claim annotations---no such gold-standard dataset exists. Therefore, we expect precision to be underestimated. That is, valid extracted claims in our more atomic approach may not match any reference claim simply because the reference set is an output of a less granular approach. Recall represents a more realistic metric of comparison, as we intend for our approach to provide coverage beyond the existing methods of claim extraction. 
}

\add{\subsection{Validation Results}}
\add{
On SciClaimHunt, our pipeline achieved precision of 0.69, recall of 0.62, and F1 of 0.65. On BioClaimDetect, we observed precision of 0.63, recall of 0.88, and F1 of 0.73.

Our pipeline achieved high recall on BioClaimDetect (0.88), successfully recovering most human-annotated claims. The lower precision (0.63) reflects atomic decomposition: our pipeline extracts multiple fine-grained claims from sentences that annotators labeled as single claims. Manual inspection confirmed that most unmatched extracted claims were valid claims absent from the reference annotations rather than extraction errors. On SciClaimHunt, we observed higher precision (0.69) but moderate recall (0.62). The lower recall likely reflects noise in the SciClaimHunt reference set, which was generated by Llama-2-13B rather than human annotators. We developed our extraction approach using SciClaimHunt examples, so we included BioClaimDetect to test robustness on an out-of-distribution dataset that uses human annotations. The stronger performance on BioClaimDetect suggests our pipeline generalizes beyond its development dataset and aligns well with human judgment of what constitutes a claim.}

\section{User Study Design}
We evaluated the efficacy of our argument-based source provenance system by comparing its use to a baseline interface \add{that represented current LLM design for QA.} This was done via a within-subjects user study with people in research-oriented roles in an organization (N=26). Participants experienced both interfaces across two tasks: a multi-paper synthesis (Task 1) and devil's advocate paper review (Task 2). The tasks were presented in a fixed order to control for task complexity progression, cognitive fatigue, and learning effects; interface condition order was randomized and counterbalanced across participants. Our study was classified as exempt from review by our ethics review board, and we pre-registered our hypotheses and methods on AsPredicted.\footnote{\url{https://aspredicted.org/wp22-d58z.pdf}}\looseness=-1

\subsection{Baseline Interface}
The baseline interface differs from \system primarily in how provenance information is presented in the \RightPanel. While \system decomposes answers into discrete claims with matched evidence from source papers, the baseline interface uses a source citation approach that mirrors how commercial LLMs currently indicate provenance. In the baseline, when users click on \SourceHighlights in the LLM's answer, the corresponding verbatim \PaperSource text appears in the \RightPanel. \add{Given the prominent use of LLMs for scholarly QA now, with source links being the main form of explanatory information, our comparison to this baseline helps us measure if scholarly LLM use can be made more deliberate and tempered by introducing provenance details.}\looseness=-1

\subsection{Task Design}
\subsubsection{Scholarly Writing Tasks}
We designed two tasks that reflect common scholarly activities researchers encounter when engaging with literature. Each task requires participants to edit LLM-generated text, while using our scholarly QA system for verification and improvement. The initial texts were generated using Google's Gemini model to ensure realistic outputs with strengths and limitations characteristic of commercial LLMs. Moreover, this same model is used for LLM-based QA, maintaining consistency in content.\looseness=-1 

\textbf{Task 1: Multi-Paper Synthesis.} This task represents the process of comparing sources for the purpose of literature review. Participants are asked to edit an approximately 300 word LLM-generated draft statement that attempts to synthesize methods and findings from all four papers. The LLM-generated synthesis naturally exhibits common LLM characteristics, including potentially incomplete integration across papers, varying levels of technical detail, and possible gaps in comparative analysis.

\textbf{Task 2: Devil's Advocate.} This task simulates a pre-submission manuscript review, where researchers must anticipate and address potential reviewer critiques. Participants are asked to edit an LLM-generated draft defense of a paper
positioned as their own manuscript, with the other three papers representing prior work.
The 300-word LLM-generated statement defending the paper's novelty and soundness was a placeholder for the argument that this task required participants to verify and edit.\hfill 
\vspace{0.4pc}

\noindent These tasks are informed by prior work categorizing how LLMs are used in scholarly settings. A recent survey study found that 81\% of 816 researchers across multiple domains already incorporate LLMs in their research workflows, with literature review being among the most frequent use cases, aligning with our multi-paper synthesis task \cite{liao2024llmsresearchtoolslarge}. The second task (devil's advocate) reflects practices where researchers use LLMs to get critical feedback on their work and examine logical consistency in their manuscripts \cite[p. 9]{10.1145/3711000}.

\subsubsection{Source Document Corpus}
Next, we describe our document selection process for the scholarly QA tasks outlined above. We chose four peer-reviewed papers on the topic of Mars exploration and planetary surface operations to serve as our corpus of source papers that the participants must ask questions about. These papers were chosen according to several criteria. The topic needed to be broad enough to be accessible across many of the research disciplines at the organization we recruited from, as well as interdisciplinary, to introduce unfamiliar material across participants. Our selected research topic involves multiple domains in engineering, science, and human factors; it is also a topic of contemporary interest~\cite{Shindell2023,lambright2014mars}, which we hoped would make the task more compelling to participants. The topics and themes needed to be coherent across papers to enable synthesis tasks, while methodologies needed to be diverse to support critical comparison across papers. \looseness=-1

We achieved this by first selecting two \add{anchor papers based on publication in peer-reviewed venues recognized within the research area. These were recommended by two domain experts at our organization who confirmed the papers represent methodologically sound contributions. These papers had no authorship or acknowledged affiliation with our organization, to minimize the likelihood that participants had prior familiarity with them. From these anchor papers, we selected two additional papers from their reference lists that addressed the same topic but used different methodological approaches to ensure diversity in contribution types.} Our final list was curated in collaboration with experts in that research area. The resulting corpus consists of four papers that look at Mars exploration architectures from different perspectives and scales. ``An Optimization Framework for Global Planetary Surface Exploration Campaigns'' \cite{ahn2008optimization} presents an optimization framework that addresses the problem of selecting landing sites, routing decisions, and maximizing scientific return under resource constraints. ``Resilient Architecture Pathways to Establish and Operate a Pioneering Base on Mars'' \cite{8396506} describes an architecture for establishing a Mars base supporting 50 people, including mission timelines, system requirements, and cost estimates. ``Pale Red Dot: a Large, Robust Architecture for Human Settlements on Mars'' \cite{doi:10.2514/6.2023-4776} proposes a Mars settlement architecture for 36 crew members distributed across two villages. And ``Leveraging Economies of Scale and Gains from Specialization for Robust Crewed Mars Architectures'' \cite{10521341} analyzes Mars missions with crew sizes from 4 to 63 members using a modeling approach that includes economies of scale and specialization effects. The papers were between 11 and 16 pages excluding references and appendices, and contained artifacts typical of scholarly papers and technical reports such as mathematical expressions, figures, and tables of quantitative information. 

\subsubsection{Question Bank}
To reduce task burden and add some standardized interaction potential, we developed a question bank of pre-written questions for each task. We pre-generated answers to these questions, enabling system responses with no lag for provenance annotations. Participants could also write their own questions.\looseness=-1

We grounded the development of these questions in the QASA framework \cite{10.5555/3618408.3619195}, which categorizes questions that scholars write about papers into three types: \textbf{surface questions}, which ``aim to verify and understand basic concepts in the content;'' \textbf{testing questions}, which focus on ``meaning-making and forming alignment with readers' prior knowledge;'' and \textbf{deep questions}, which ``ask about the connections among the concepts in the content and elicit advanced reasoning.'' Each type of questions is further categorized in subtypes with examples provided in the QASA paper.\looseness=-1

We used Gemini to generate the question bank in the form of 1-2 questions for each relevant subtype. The prompt for Gemini included: (1) the QASA question type definitions, subtypes, and examples; (2) our task descriptions; and (3) the pre-generated texts that participants would edit. For Task 1 (Multi-Paper Synthesis), we used questions from the \textbf{testing} category: examples, quantitative comparisons, definitions, and compare/contrast questions. These question types support cross-paper synthesis by prompting participants to align information and make connections across sources. For Task 2 (Devil's Advocate), we used questions from the \textbf{deep} category: causal relationships, goals and motivations, procedural details, rationales, and expectations. These question types mirror how critical reviewers interrogate a manuscript's argumentation and methodology. We did not pre-write \textbf{surface} questions because they trigger basic fact-lookup that participants can easily perform on their own, and are misaligned with our aim to elicit the meaning-making and higher-order reasoning required by synthesis and critique.\looseness=-1

\subsection{Procedure and Flow}
Participants completed a pre-study survey to indicate interest, share demographic information, and rate their familiarity with the source document corpus (too much familiarity was used as an exclusion criteria; Section~\ref{sec:participants}). Sessions were then conducted remotely and asynchronously via a web browser, and included the following steps:\looseness=-1
\begin{itemize}
    \item Participants land on an About page that summarizes the study, provides them with contact information for the research team, and lists the study steps and their approximate durations. This page informs them that they may choose not to participate at any time and gathers consent. 
    \item Next, participants complete the Trust in Explainable AI (TXAI) survey instrument \cite{10.3389/fcomp.2023.1096257}. Following \citet{10.1145/3544549.3585808}'s recommendation, we exclude the reverse-coded question (item 6: ``I am wary of the AI''), as doing so improves internal consistency. They are asked to think of an LLM-based system that they have used recently for scholarly tasks and to answer the questions specifically with that system in mind. 
    \item After completion of the initial trust survey, participants view a tutorial that illustrates the interface they will use for Task 1.\looseness=-1
    \item After viewing the tutorial, the participant is brought to the interface they will use for Task 1. This page includes the task environment, the chatbot, and the intervention specific to the current condition. The participant uses the interface and edits the drafted text until they are satisfied with the result. Our guidance was to aim for 20 mins to complete the task. 
    \item After submission of the edited text, the participant takes four post-task surveys: 
    (1) the TXAI scale \cite{10.3389/fcomp.2023.1096257}, only being asked in the context of the LLM used, not the entire interface (\add{with instructions and edits to TXAI scale items to clarify this}); (2) a two-item confidence assessment, where they indicate on a scale of 1 to 7 their confidence in the edits they made (``I am confident in the edits I made to the text'') and in the final text (``I am confident in the quality of the final text''); (3) the NASA-TLX scale to measure workload \cite{HART1988139}; and (4) the SUPR-Q scale to evaluate the usability of the application \cite{sauro2015supr}.\looseness=-1
    \item After completion of the post-task surveys, the participant is taken to a tutorial for the interface used for Task 2. They then perform the second task and complete the post-task surveys again with Task 2 details in mind.
    \item After completion of the second post-task surveys, the participants land on a final set of questions that ask them to indicate which system they preferred, and provide an optional free-response box for feedback on the systems. This is followed by a debrief page that explains the study in more detail. \looseness=-1
\end{itemize}

\subsection{Participants}~\label{sec:participants}
We recruited people in research-oriented roles (both scientists and research engineers) across National Aeronautics and Space Administration centers---they were asked to express interest in participating in our study by completing a pre-screen survey. Participation in this study was voluntary, i.e., there was no monetary incentive provided for completing the study. 78 people completed our pre-screen, of which 74 met our inclusion criteria. We excluded participants who rated their familiarity with any of the papers in the corpus as greater than 3 out of 7, as these people would have an unfair advantage given their prior knowledge of the work. Of these 74 people who met our inclusion criteria, 38 ultimately participated in our study. The remaining 36 did not complete the study due to scheduling conflicts, time constraints, or lack of response to follow-up communications.

We report results from 26 participants after excluding data from 12 who did complete the study but their data was not valid. These 12 participants were excluded for the following reasons: (1) they took less than 10 mins to complete each task, without any interaction with the components of our systems; (2) they added gibberish text in their responses; and (3) their qualitative responses indicated that system latency had prevented them from engaging with the information (caused by load balancing issues in the backend). This exclusion process was conducted based only on feedback and task duration, without referencing our primary outcome measures, to avoid biasing the results. 

The sample included 20 men and 6 women. Most participants (24 of 26) were between the ages of 25 and 54. 
The majority were highly experienced, with 18 participants having between 6 and 20 years of professional experience, and all but two held an advanced degree (13 Master's, 11 Doctoral). Regarding their familiarity with LLMs, 21 participants reported using LLM-based tools either daily or weekly. \add{All participants had expertise in research and engineering in the sciences, including domains like chemistry, physics, materials science, and aerospace.}

\subsection{Measures}
We use three primary measures to understand the outcomes of using \system compared to the baseline interface for scholarly tasks, as well as additional exploratory measures to understand participants' experiences with our system.

\subsubsection{Primary Measures}
Our system is intended to help people appropriately use LLMs for scholarly tasks. We measure aspects of this usage via three primary metrics: people's trust in LLMs after using the interfaces, their reliance on the placeholder LLM-generated text for the scholarly tasks, and their confidence in the LLM and their output.\looseness=-1

\textbf{Trust. }We measure subjective trust using the validated TXAI scale \cite{10.3389/fcomp.2023.1096257} at three points: baseline (pre-study) and after each task. Items are rated on a 1-7 Likert scale and averaged. We use the pre-study trust value descriptively for contextualizing initial behaviors; only the trust values for individual conditions are used for comparisons. \add{The scale items are asked in the context of the LLM used for the Q\&A, rather than the other design features of the interfaces.}

\textbf{Reliance. }We operationalize reliance based on changes to the LLM-generated placeholder text provided for the scholarly task. We calculate a token-level edit similarity based on Levenshtein distance \cite{1966SPhD...10..707L} between the original LLM-generated draft ($x$) and the participant's final edited text ($y$), normalized by the token count of the longer token sequence:\looseness=-1
\begin{equation}
    \mathrm{Reliance} \;=\; 1 \;-\; \frac{\mathrm{LD}\!\big(W(x),\,W(y)\big)}{\max\{\lvert W(x)\rvert,\,\lvert W(y)\rvert\}} \,
\end{equation}
where $\mathrm{LD}$ is Levenshtein distance over tokens and $W(\cdot)$ maps a string to a sequence of lowercased lemmas after removing standard English stop words while retaining negations (\emph{no, not, nor, n’t}) to preserve polarity.\footnote{Lemmatization collapses inflectional variants (e.g., \emph{optimize/optimized/optimization}), and stop-word removal focuses the metric on content-bearing terms; operating at the word level aligns the unit of comparison with typical revision actions and reduces sensitivity to punctuation and formatting noise.} Normalizing by $\max\{\lvert W(x)\rvert,\,\lvert W(y)\rvert\}$ ensures large expansions or pruning register as reduced reliance. Values for reliance range from $[0,1]$, with $1$ indicating no edits (full reliance) and lower values reflecting greater rewriting. \add{We chose token-level Levenshtein distance as our reliance measure because our research question concerns behavioral engagement with the text---specifically, whether participants physically edited the LLM output---rather than the semantic quality of those edits. Levenshtein distance directly captures editing actions: additions, deletions, and substitutions that participants made to the draft at the word-level, excluding minor edits (e.g., to stop words). 
Our measure treats all edit operations equally regardless of their semantic impact, which aligns with our goal of understanding whether provenance information changes low-level editing behavior.}

\textbf{Confidence. }We measure participants' subjective confidence in each task output on two 7-point items (``I am confident in the edits I made to the text'' and ``I am confident in the quality of the final text.'' We average the two items to form a per-condition confidence score given internal consistency in the measures, calculated as Cronbach’s $\alpha=0.89$, 95\% CI = [0.81,0.93].

\subsubsection{Secondary Measures}
We collect additional subjective and behavioral measures to contextualize primary outcomes. \textbf{Workload} is assessed post-task for each interface using the NASA-TLX questionnaire \cite{HART1988139}; we compute the raw (unweighted) overall score as the mean of the five subscales (omitting the second item, ``Physical Demand''). \textbf{Perceived usability} is assessed post-task with the SUPR-Q questionnaire \cite{sauro2015supr} and averaged. \textbf{Interaction logs} include time-on-task; number of chat questions; and clicks on provenance features (e.g., source cards, claim/evidence items, focus toggles). We also record \textbf{interface preference} and \textbf{open-ended feedback} at the end of the study.\looseness=-1
\begin{table*}[ht!]

\label{tab:results_summary}
\begin{tabular}{rcccc}
\toprule
\textbf{Measure} & \textbf{Baseline Mean} & \textbf{\system Mean} & \textbf{Statistic} & \textbf{Effect Size} \\
\midrule
\multicolumn{5}{c}{Primary Factors} \\\midrule
Trust & 4.22 $\pm1.22$ & 3.68 $\pm1.24$  & $t = 2.61,\ p=.015\ast$ & 0.44 \\
Reliance & 0.73 $\pm0.21$ & 0.75 $\pm0.29$ & $W = 146.00,\ p=.313$ & -0.23 \\
Confidence & 4.27 $\pm1.51$ & 4.05 $\pm1.58$ & $t = 0.64,\ p=.525$ & 0.14 \\\midrule
\multicolumn{5}{c}{Experiental Factors} \\\midrule
Cognitive Load & 4.10 $\pm0.83$ & 4.12 $\pm0.78$ & $t = -0.21,\ p=0.838$ & 0.03 \\
Usability & 5.05 $\pm1.08$ & 4.48 $\pm1.17$  & $W = 71.50,\ p=.026\ast$ & 0.52 \\
Clicks & 8.26 $\pm6.02$ & 12.47 $\pm14.96$  & $W = 97.50,\ p=.137$ & -0.35 \\
Messages & 3.67 $\pm2.57$ & 4.08 $\pm2.84$ & $t = -0.79,\ p=.438$ & 0.16 \\\midrule
\multicolumn{5}{l}{$t$ = Paired t-test, $W$ = Wilcoxon signed-rank test, $^{***}p < .001$, $^{**}p < .01$, $^{*}p < .05$, $^{\boldsymbol{\cdot}}p = .1$} \\\bottomrule
\end{tabular}
\caption{Comparison of measures between baseline and \system conditions.}
\label{tab:analysis}
\end{table*}

\subsection{Analyses}
We analyze our primary dependent variables using paired comparison tests between the baseline and \system data: t-tests when the distribution is normal, Wilcoxon signed-rank tests otherwise. Normality of the data is tested using the Shapiro-Wilk test. All statistical tests are two-tailed, with an alpha level $0.05$. We also report comparison testing outputs for our exploratory experiential variables: cognitive load, usability, clicks, and messages; these are intended for descriptive purposes and to understand the differences in our primary measures. We also report the final interface preference as a frequency count. The qualitative data is coded using \citet{Braun01012006}'s inductive approach, with the intention to understand specific outcomes for what people liked, disliked, and wanted to add to our system design. We conduct Spearman correlation analyses between demographic, experiential, and primary outcome variables to understand relationships between measures and how they differ across interface conditions.

\section{Results}
This section presents the results of our within-subjects study. We first report results from comparisons between \system and baseline on our primary outcome and secondary experiential measures, then correlation analysis and qualitative themes to contextualize these numbers. Table \ref{tab:analysis} presents an overview of the descriptive and comparison outputs for all our measures. 

\subsection{Primary Measures: Trust, Reliance, and Confidence}\label{sec:primary-results}
A paired t-test showed a statistically significant effect of the interface on subjective trust \add{on the LLM}. Participants reported significantly lower trust in LLMs when using \system compared to the baseline ($t(25)=2.61,p=0.015)$), with a medium effect size (Cohen's $d=0.44$). This supports our hypothesis that claim-evidence annotations encourage more caution towards LLM use in scholarly settings. However, despite the reduction in trust, we found no significant difference in behavioral reliance between the two conditions ($W=146,p=0.313$); nor in self-reported confidence ($t(25)=0.64, p=.525$).

We propose three potential explanations for the discrepancy between the drop in trust and the non-significant change in reliance behavior for further consideration. First, the cognitive effort required to learn and navigate the novel \system interface within the limited time may have diverted participants' attention from the primary writing task, reducing the likelihood of extensive edits. Second, while \system's granular details made participants more skeptical of the LLM itself, the system's transparent design may have been perceived as more trustworthy overall, shaping reliance in a way that counteracted their caution towards the LLM. Finally, the interface may have had a bimodal effect, where it increased trust for some users who valued the verification features while decreasing it for others who were confronted with the LLM errors. However, coupled with the limited editing behavior of several participants given time constraints, this bimodal effect is lost in regression to the mean. We unpack these further in our qualitative findings below.\looseness=-1

\subsection{Secondary Experiential Measures}
Our results show that the increased critical scrutiny afforded by \system came at the cost of lower perceived usability. A Wilcoxon signed-rank test indicated that the \system interface was rated as significantly less usable than the baseline ($W=71.5,p=0.026$), with a medium effect size ($r=0.52$). We consider two possible explanations for these values. First, scholarly synthesis is an inherently demanding task, and \add{this intrinsic difficulty may have been exacerbated by the feature-richness of \system. Indeed, prior work has often found this to be the case with rich explanatory outputs~\cite{kaur2024interpretability,10.1145/3411764.3445717,rastogi2022deciding}.} Second, our lab study did not afford participants enough time to move past the initial learning curve associated with the novel interface. In a real-world field deployment where users could develop expertise in the system over time, perceptions of usability might be different.


\add{Click counts showed a marginally significant difference between conditions} ($W=97.50,p=0.137$), with participants clicking more in \system ($M=12.47,SD=14.96$) than in the baseline ($M=8.26,SD=6.02$). \add{However, higher click counts are ambiguous as an engagement indicator. More clicks could reflect deeper exploration of provenance information (the intended use), but could also indicate: interface inefficiency requiring more actions to accomplish equivalent goals; confusion leading to exploratory clicking; or repeated attempts to understand unfamiliar features. The large standard deviation in \system clicks suggests highly variable engagement patterns across participants. Without click sequence analysis or qualitative observation of navigation patterns, we cannot say whether the additional clicks represent productive verification behavior or interaction overhead. We therefore interpret this finding cautiously.}
We return to this tradeoff between complex information presentation and usability in the qualitative results and Discussion. Other experiential measures---cognitive load and number of messages sent to the LLM---were similar across the two interfaces.\looseness=-1

\begin{table*}[th!]
\centering
\resizebox{0.98\linewidth}{!}{
\small
\begin{tabular}{rlllllllllll}
  \toprule
 & Ed. & Exp. & Freq. & Gender & Conf. & Rel. & Trust & Cog. Load & Usability & Clicks & Messages \\ 
  \midrule
  \textbf{Demographics} & & & & & & & & & & & \\
Age  &  0.47* &  0.79*** &  0.19 & -0.04 & -0.03 &  0.12 & -0.08 & -0.29 & -0.08 & -0.09 & -0.02 \\ 
  Education  &  &  0.53** &  0.10 & -0.16 &  0.07 &  0.18 &  0.02 & -0.11 & -0.05 & -0.07 &  0.16 \\ 
  Experience  &  &  &  0.10 & -0.20 & -0.01 &  0.02 & -0.01 & -0.21 & -0.10 &  0.04 &  0.14 \\ 
  Frequency  &  &  &  &  0.06 &  0.05 &  0.13 &  0.00 &  0.27 &  0.06 & -0.06 &  0.27 \\ 
  Gender  &  &  &  &  & -0.07 &  0.34 & -0.10 &  0.16 &  0.03 &  0.13 &  0.05 \\ 
  \textbf{Primary} & & & & & & & & & & & \\
  Confidence   &  &  &  &  &  &  0.24 &  0.60** & -0.08 &  0.47* & -0.13 &  0.07 \\ 
  Reliance  &  &  &  &  &  &  &  0.57** &  0.21 &  0.50* &  0.02 &  0.11 \\ 
  Trust   &  &  &  &  &  &  &  &  0.31 &  0.74*** &  0.11 &  0.30 \\ 
  \textbf{Experiential} & & & & & & & & & & & \\
  Cognitive Load  &  &  &  &  &  &  &  &  &  0.29 &  0.26 &  0.47* \\ 
  Usability  &  &  &  &  &  &  &  &  &  & -0.10 &  0.22 \\ 
  Clicks  &  &  &  &  &  &  &  &  &  &  &  0.16 \\ 
   \midrule
\multicolumn{12}{l}{$^{***}p < .001$, $^{**}p < .01$, $^{*}p < .05$} \\\bottomrule\end{tabular}
}
\caption{Spearman correlations in the baseline condition. Education refers to level of educational attainment, Experience refers to years of experience, Frequency refers to frequency of AI use, and Clicks and Messages refer to the total counts of each.} 
\label{tab:corr_baseline}
\end{table*}

\begin{table*}[th!]
\centering
\resizebox{0.98\linewidth}{!}{
\small
\begin{tabular}{rlllllllllll}
  \toprule
 & Ed. & Exp. & Freq. & Gender & Conf. & Rel. & Trust & Cog. Load & Usability & Clicks & Messages \\ 
  \midrule
  \textbf{Demographics} & & & & & & & & & & & \\
Age &  0.47* &  0.79*** &  0.19 & -0.04 & -0.20 &  0.00 &  0.03 & -0.18 &  0.23 & -0.38 & -0.28 \\ 
  Education  &  &  0.53** &  0.10 & -0.16 & -0.41* &  0.25 &  0.01 & -0.15 & -0.08 &  0.04 & -0.24 \\ 
  Experience  &  &  &  0.10 & -0.20 & -0.29 &  0.22 & -0.01 & -0.12 &  0.07 & -0.20 & -0.23 \\ 
  Frequency  &  &  &  &  0.06 & -0.19 &  0.11 & -0.12 &  0.17 & -0.09 & -0.11 &  0.08 \\ 
  Gender  &  &  &  &  &  0.03 &  0.14 & -0.29 &  0.13 &  0.05 & -0.28 & -0.11 \\ 
  \textbf{Primary} & & & & & & & & & & & \\
  Confidence  &  &  &  &  &  & -0.06 &  0.61*** & -0.28 &  0.62*** &  0.08 & -0.13 \\ 
  Reliance  &  &  &  &  &  &  &  0.06 & -0.11 &  0.16 & -0.11 & -0.32 \\ 
  Trust  &  &  &  &  &  &  &  & -0.27 &  0.73*** & -0.03 & -0.11 \\ 
  \textbf{Experiential} & & & & & & & & & & & \\
  Cognitive Load  &  &  &  &  &  &  &  &  & -0.28 &  0.22 &  0.04 \\ 
  Usability  &  &  &  &  &  &  &  &  &  & -0.22 & -0.32 \\ 
  Clicks  &  &  &  &  &  &  &  &  &  &  &  0.32 \\ 
   \midrule
\multicolumn{12}{l}{$^{***}p < .001$, $^{**}p < .01$, $^{*}p < .05$} \\\bottomrule \end{tabular}}
\caption{Spearman correlations in the \system condition. Education refers to level of educational attainment, Experience refers to years of experience, Frequency refers to frequency of AI use, and Clicks and Messages refer to the total counts of each.} 
\label{tab:corr_papertrail}
\end{table*}

\subsection{Correlation Analysis}
To better understand some of our findings above, we used Spearman's coefficient to explore relationships between demographic, experiential, and primary outcome variables for the baseline (Table \ref{tab:corr_baseline}) and \system (Table \ref{tab:corr_papertrail}).

One difference between the interfaces is the relationship between reliance and trust. In the baseline condition, reliance was significantly and positively correlated with trust ($r=0.57,p<0.01$) and usability ($r=0.60,p<0.05$). This suggests that the participants who trusted the LLM more and found the system more usable also tended to rely on the LLM output more heavily (i.e., edited the text less). In contrast, these correlations disappear in the \system condition. While the strong relationships between trust, confidence, and usability remained, none of these factors were significantly correlated with how much a participant chose to edit the LLM output. This suggests that the claim-evidence features in \system may have decoupled the simple relationship between a user's general trust in LLMs and their reliance on LLM-generated text. We suspect this is due to the same bimodal shift in behavior change as a consequence of reduced trust that is described in Section~\ref{sec:primary-results}.

\add{Additionally, we observed a significant negative correlation between education and confidence with \system ($r=-0.41, p<.05$) that was absent in the baseline. This suggests that more highly educated participants may have been more sensitive to the provenance information shown by \system, leading to reduced confidence in their outputs. We did not find any other significant correlations between participant demographics and our measures.\looseness=-1}

\subsection{Qualitative Findings}
Our qualitative analysis of participants' study feedback revealed three primary themes that help explain the quantitative results: external constraints that shaped reliance behaviors; tensions between information richness and usability; and paradoxical trust behaviors despite recognition of verification needs.

\subsubsection{Factors Affecting Reliance Behaviors}
\textbf{Time Pressure and System Performance.} The most frequently cited barrier to meaningful engagement was time constraints, mentioned explicitly by over half of the participants. Guiding participants to complete the tasks in around twenty minutes meant they had insufficient time to deeply engage with the provenance features or verify claims against source documents. This constraint was exacerbated by system latency, with participants describing response times as ``excruciatingly slow,'' (P1) and ``much slower than others [LLMs] that I have used'' (P3). This led to many participants having a similar experience to participant 23, who ``didn't make any edits to the text,'' and ``[instead] asked a few questions to query and verify the accuracy of the responses.'' The combination of limited time and slow responses impacted how participants engaged with the provenance features, and participants noted this explicitly as well, like P15: ``I feel the time constraints fundamentally changed the way I used and trust the AI output. I depended on the AI more because of the short time constraint and ultimately spent less time revising and checking the results than I otherwise would have.'' 

\textbf{Task Authenticity and Engagement Strategies.} Participants approached the tasks with divergent strategies based on their framing of the study context. Some experts wanted to at least skim the papers on their own first before doing the task, and consequently expressed frustration with the artificial nature of evaluating less familiar papers, though in their research domain: ``To produce more informative results, it would've been necessary for the participants to either familiarize themselves with the four papers before the test or to supply their own four papers with which they are thoroughly familiar'' (P5). Conversely, some participants explicitly acknowledged the artificial nature of the study context and calibrated their engagement accordingly. As P2 explained, ``I did not have to actually read any of the papers. I guess that means I trusted the AI tool to be accurate. I didn't believe everything it was telling me was accurate, but I figured it was close enough to complete the task at hand.'' This divergence in engagement strategies between those seeking ecological validity and those accepting the study's limitations likely influenced the variation in reliance scores, as participants' editing behaviors reflected their differing interpretations of what the task demanded. \looseness=-1

\subsubsection{The Information Complexity--Usability Tradeoff}
Participants consistently recognized the need for access to complex information for scholarly tasks while struggling with its presentation. The quality of LLM outputs was frequently criticized, with one participant comparing it to ``an eighth grader'' (P25) and another to ``what I would expect from a novice researcher placed in the same bind'' (P4). Given their dissatisfaction with the LLM output quality, many participants expressed a desire for deeper engagement with the source materials, suggesting they would have preferred to read all papers thoroughly before attempting the tasks. Yet this desire for comprehensive review reflects an underlying tension in scholarly AI tool evaluation. While thorough paper familiarity might improve task performance in a study setting, such exhaustive pre-reading is specifically what these tools aim to help researchers avoid in practice, where the volume of literature makes reading every paper impractical. Overall, this dissatisfaction drove appreciation for the detailed provenance features, but the implementation created significant usability challenges. 

\textbf{Interface Complexity and Physical Constraints.} Given their desire for richer, nuanced outputs, participants appreciated the theoretical value of the argument-based provenance information. In fact, we received some invitations to share our system and findings more broadly in the organization, noted by participants in their feedback responses to the study. However, the implementation revealed gaps between our design goals and user experience. 

Our design goal DG2 aimed to ``minimize interaction overhead while preserving exploration depth,'' yet participants found the interface to be ``cluttered...for a fairly small laptop screen'' (P25), with nested windows that allowed reading ``only a line or two at a time.'' This directly contradicted our intention to optimize space/time resources. Similarly, while DG3 sought to ``enable flexible verification workflows,'' the rigid presentation of claim-evidence cards actually constrained participants' verification strategies. The tension between appreciation and frustration can be seen in P24's feedback, who found the interface was ``interesting and [having] value,'' but that it needed to ``communicate this value more easily.'' The question bank partially addressed DG1's goal of ``graduated cognitive engagement'' by providing an entry point for exploration. As P18 noted:  ``I could not have done this without the suggested questions.'' However, this single success could not overcome the broader failure to achieve balance between complexity and usability. \add{We consider this usability feedback in our Discussion of future design implications.}\looseness=-1

\subsubsection{The Ethos of Grounding LLM Outputs in Provenance Information}
Despite usability frustrations, most participants understood and valued the project's underlying goals. While there were exceptions (``this interface/model completely misses how I use an LLM for research or paper writing purposes'' (P21)), the majority recognized the critical need for verification capabilities in scholarly LLM tools. Participants articulated various aspects of this need: the importance of ``fact checking'' (P1); ``trying to figure out what was missing or possibly incorrect'' (P2); and evaluating ``how credible or trustworthy the application's responses were'' (P18). P6 framed it as an ethical concern, warning against ``the temptation to use AI to write material for you,'' which they equated with ``a general dumbing down of the world.''

While usability issues limited positive experiences---only a small subset, like P3, ``found the tasks easy to complete with the LLM QA interface provided''---participants nonetheless appreciated the conceptual value of argument-based provenance. P14's feedback articulated precisely what \system aimed to achieve:

\begin{quote}
    If more tools could be given to find the location of specific claims within the papers, that would be helpful. I find that errors often occur in LLMs through stripping context. To help with overall accuracy, every effort should be made to help the human user track down the original context to verify claims.
\end{quote}
This recognition that LLM errors stem from ``stripping context'' and that verification requires tracing claims to their original sources supports the premise of claim-evidence provenance that we prioritized in \system.

\section{Discussion}
Our evaluation of \system shows that while argument-based provenance can encourage healthy skepticism toward LLM outputs, translating this into changed reliance behavior requires overcoming substantial barriers related to time, usability, and ingrained patterns of tool use. The relationship between attitudes and actions in human-AI collaboration is complex, particularly in time-constrained, cognitively demanding contexts like scholarly tasks. Significant usability costs of our implementation likely contributed to this trust-behavior gap.
\add{Below, we first contextualize our findings within prior work on trust and reliance for AI-assisted decision-making. Grounded in that argument, we present theoretical and design implications towards high-utility paths forward for our new scholarly AI assistance setting; and end with computational opportunities for improvement.}

\subsection{Trust-Behavior Gap: Why Didn't Lower Trust Change Reliance?}
\add{While our methodological and implementation constraints likely played a role in the trust-behavior change gap (details in Limitations below), our methods and findings share similarities with prior work on explainable AI (XAI) and AI-assisted decision making. These prior work contexts similarly show that explanations can increase reliance on both correct and incorrect predictions~\cite{lai2021towards,10.1145/3706598.3714020}, through under certain conditions they can reduce overreliance~\cite{10.1145/3579605}; enforced engagement with them can reduce user satisfaction~\cite{10.1145/3449287,de2025cognitive}; and their presentation does not often cultivate behavior change as people continue to defer to system outputs under time pressure or cognitive load~\cite{kaur2024interpretability,rastogi2022deciding,abdul2020cogam}. We hypothesize three reasons for the consistent results across these contexts and ours.}\looseness=-1

\add{\textit{First}, people default }to System 1 thinking---fast, automatic, and heuristic-based reasoning---even when they intellectually recognize the need for deliberation. \add{The Dual Process theory of cognition distinguishes between System 1's efficient but error-prone processing and System 2's effortful analytical reasoning~\cite{evans2013dual,kahneman2011thinking}; also termed ``bounded rationality''~\cite{simon1997models}. As XAI work has shown, making AI systems more explainable can sometimes exacerbate this problem by providing convenient narratives and analogic thinking devices that feel like comprehension without requiring genuine verification~\cite{danry2023don,kaur2024interpretability}. Similarly, our claim-evidence interface attempted to engage System 2 thinking by requiring users to evaluate logical connections between claims and sources}, but 
the cognitive cost of verification remained too high relative to the perceived benefits within the constrained study context. Prior work has shown that complementary human-AI team performance depends on accurate mental models of AI capabilities rather than trust alone~\cite{Bansal2019BeyondAT}, suggesting that a single session may have been insufficient for participants to develop the mental models needed to act on their skepticism.

\add{\textit{Second}, the intrinsic incentives that drive people away from System 1 thinking are missing in AI-assisted settings.} \citet{klein2009}'s reconciliation of naturalistic decision-making vs. heuristics-based outcomes identifies two necessary conditions for genuine intuitive expertise: high-validity environments (with stable regularities to learn) and opportunities to learn these regularities through feedback. AI-assisted settings may violate both conditions. In our case, LLMs produce errors unpredictably---plausible-sounding synthesis that may contain subtle omissions, over-generalizations, or unsupported leaps---making it difficult to develop stable heuristics for spotting problems. Moreover, people receive limited feedback on whether their edits successfully improved accuracy. Without this feedback loop, people cannot learn when their skepticism should translate into action versus when selective reliance was appropriate, giving them no motivation to engage deliberately.\looseness=-1

\add{\textit{Third}, designing systems that successfully shift cognitive behavior is a hard problem. \system represents one approach---introducing what has been called ``design friction''~\cite{kaur2024interpretability,naiseh2021nudging}, ``seamful design''~\cite{10.1145/3531146.3533135,ehsan2024seamful} or a ``cognitive forcing function''~\cite{10.1145/3449287} in XAI work to scholarly contexts, by requiring that people engage with claim-evidence structures before accepting LLM outputs. However, as both prior work and our findings demonstrate, such interventions face a difficult balancing act. If the friction is too burdensome, it hurts usability and people circumvent or abandon the system entirely~\cite{kaur2024interpretability,sanneman2024information,gaube2024underreliance}; our study feedback suggests we erred in this direction. If the intervention is too lightweight, it fails to engage System 2 thinking and people maintain their original behaviors. Prior work has even framed this as a cost-benefit analysis, seeking to understand when to prioritize different types of thinking: a challenge that persists~\cite{rastogi2022deciding}. }

\add{If we anticipate some continued consistency in these two types of AI-assisted settings (prior work on XAI and AI-assisted decision making and our scholarly context), we can ground our implications in what would be different from the ideas explored before. While there remains tremendous potential for translating successes across these two contexts, we consider how to make progress on what is uniquely challenging for scholarly QA and writing via theoretical, design, and computational implications below. }

\subsection{\add{Theoretical Implication: From Trust to Trustworthiness}}
What does it mean to study appropriate trust and reliance on LLMs? Is the goal to always lower trust? As we note above, this would not be ideal from a cognitive standpoint, and people might become so skeptical that they under-utilize a system. Appropriate trust is particularly challenging in a setting when the model is inherently black-box with no faithful representation of interpretability. We consider an alternate framing: making an LLM trustworthy rather than changing user trust.\looseness=-1 

\subsubsection{Argument-Grounded Provenance as a Metric of LLM Trustworthiness}
The backend architecture of \system can be extended beyond a user-facing tool into a formal trustworthiness benchmark for scholarly QA systems. Our method of using both local (claim-level matching) and global (claim coverage) information structures is a detailed way to evaluate LLM outputs that goes beyond surface-level metrics. This approach complements expert-derived error schemas \cite{10.1145/3772318.3791843} that categorize LLM failures in scholarly QA across dimensions including correctness, completeness, hallucinations, interpretation, and synthesis quality. Comparing answer claims against relevant source claims enables quantitative measure of both faithfulness and completeness. This approach is particularly effective for highlighting critical omissions, which is a difficult-to-detect failure mode in current systems. Such a benchmarking framework could enable systematic comparison of commercial LLMs to identify which are best suited for scholarly QA, support development of specialized scholarly LLMs through granular feedback, and allow researchers to audit LLMs before deployment.\looseness=-1
We present this argument-grounded provenance matching approach as an ``application-grounded evaluation''~\cite{doshi2017towards} for establishing LLM trustworthiness in scholarly settings, complementing existing methods while addressing the specific needs of scholarly users.

\subsection{Implications for Design}
\add{Based on design lessons from prior work and our study, we consider the following implications to address how future systems might better balance the competing demands of information completeness and interaction fluidity.}

\subsubsection{Managing the Cost of Information Granularity with Adaptive Provenance}
Our results show that detailed provenance encourages caution but can overwhelm users. Future systems should investigate adaptive provenance that dynamically adjusts detail levels based on context; for example, providing simple source links for straightforward questions but surfacing full claim-evidence interfaces for complex or controversial queries. The interface might default to simpler cues even in settings requiring greater complexity (perhaps just the claim coverage bar/indicator), and allow progressive disclosure of deeper structures on demand. Such adaptive systems would need to intelligently categorize query complexity, perhaps using the QASA framework's distinction between surface, testing, and deep questions \cite{10.5555/3618408.3619195}, or \add{borrowing from user-centered question banks of prior XAI work~\cite{10.1145/3313831.3376590,arya2019one,sipos2023identifying}.}

\subsubsection{Scaffolding Critical Thinking Through Progressive Engagement}
Claim-evidence interfaces can potentially act as effective scaffolds for critical thinking when users have sufficient cognitive resources. Scholarly information systems should progressively develop users' verification skills over time through tutorial modes that guide users through verification on simpler tasks before expecting independent critical evaluation on complex syntheses. \add{This learning-based approach has shown promise in improving appropriate reliance for AI-assisted decision making~\cite{gao2024going}.} Systems could also provide verification templates tailored to specific scholarly tasks\add{---another approach with successful artifacts in improving responsible AI practices~\cite{madaio2020co,deng2025weaudit}}. Finally, ludic design elements \cite{10.1007/978-3-540-28643-11} could reward thorough verification behaviors through what \citet{nguyen2022playfulness} calls ``epistemic playfulness,'' engaging with provenance through game-like challenges rather than purely for accuracy assessment. However, ludic elements must carefully encourage genuine exploration rather than superficial point-scoring. \hfill\newline

\noindent \add{A key difference between scholarly QA and prior work on AI-assisted decision-making is that appropriate reliance is conceptually different for these settings. Decision-making offers binary or multi-class constraints, while scholarly QA is open-ended---lacking ground truth and depending on question and user context. This necessitates combining designs from prior work with richer workflow analyses, which we hope future work will tackle.}
\subsection{Computational Considerations}
Performing deep semantic analysis across large documents in real-time for every user query is computationally expensive, leading to the high latency that hurt user experience in our study. \add{We consider some ways in which this can be improved in future iterations of our computational approach.} Our backend demonstrates load-balancing through offline preprocessing, but we retained LLM processing in real-time for answer generation and claim matching (Stages 2 and 3), creating sequential bottlenecks. Each query required multiple LLM calls that could not be parallelized due to dependencies—answers must be generated before claims can be extracted, and paper claims must be filtered before matching. Future implementations could address these through aggressive caching (pre-computing common query patterns), parallel processing (extracting claims from answer chunks simultaneously), or faster inference infrastructure. The flexible extraction approach allows strategic selection of methods: similarity-based extraction for high-volume filtering, LLM-based extraction for user-selected claims requiring deeper analysis. This could extend to federated architectures where institutions pre-process document collections offline, sharing only lightweight claim indices for real-time matching~\cite{manghi2019openaire,Bilder2015}.\looseness=-1

\section{Limitations and Future Work}
\add{\textbf{Task Realism. }This took two forms: (1) participants engaged with unfamiliar papers within artificial time constraints, rather than conducting authentic literature searches or working with materials from their own research; and (2) the 20-30 minute task duration guidance, that prevented the deep engagement that characterizes scholarly work in practice.} While this standardization was necessary for measuring reliance consistently across participants, it may have fundamentally altered engagement patterns. 
Future field studies should examine \system's effectiveness when researchers use it for their actual work, with familiar literature and self-directed questions. \add{Extensions of this experimental setting to include the writing elements of research reading (e.g., as done to generate workflow datasets in~\cite{le2025scholawrite}), also deserve more attention, but was out of scope for our user study.}

\add{\textbf{Metrics and Measurement.}} First, our operationalization of reliance through edit distance may not fully capture the nuanced ways participants engaged with LLM assistance. The measure conflates various behaviors (from wholesale acceptance to strategic delegation), and creates a potential confound. Participants might maintain high textual similarity not from overtrust in the LLM, but from trust that \system would alert them to problems requiring intervention. Future work should develop more sophisticated behavioral measures that distinguish between passive acceptance and informed delegation. \add{Relatedly, when reporting subjective trust using the TXAI scale, participants may not even have distinguished between the interface and the LLM. Component-specific trust measures will be important for future verification. Finally, we did not evaluate the quality of participants' edited texts. While our focus was to capture behavioral differences, a quality assessment would help identify 
whether unsupported claims were removed, omitted information was added, and if the overall argumentation improved. 
Future work should include blind expert evaluation of output quality to complement behavioral measures.}

\add{\textbf{Design Generalizability.} \system instantiates one approach to claim-evidence provenance within a specific interface paradigm (three-panel layout with coordinated views). Alternative designs, such as inline annotations, progressive disclosure, or conversational verification, might produce different trust-reliance dynamics. Our findings speak to the value of argument-grounded provenance as a concept but not to the optimality of our particular implementation. Similarly, the basis of this design may be approached differently. Domains within and outside of STEM may have varying argumentation conventions}, and other scholarly tasks may benefit from different provenance approaches. 
We do not capture rebuttals, warrants, backings, or qualifiers \cite{Toulmin_1958}; \add{or other argumentation styles~\cite{walton2010types}. Future work should examine how argument-based provenance may be designed across diverse academic disciplines, argumentation details, and task types. \add{Additionally, our system assumes source documents follow a clear claim-evidence structure. If source papers contain unsupported claims or lack explicit evidentiary reasoning, the extraction pipeline may produce incomplete or misleading provenance information. Users cannot easily distinguish whether an answer claim is flagged as unsupported due to a genuine LLM error or due to insufficient coverage in the source corpus—a limitation that may contribute to the lack of behavior change we observed.}\looseness=-1

\textbf{Participant Pool. } Our sample of researchers at a single organization represents a narrow slice of potential users. Scholars in different disciplines may have varying familiarity with structured argumentation (e.g., legal scholars vs. bench scientists), different verification norms, and different time pressures. Students, who are increasingly using LLMs for literature review~\cite{10.1145/3711000}, face distinct challenges around domain knowledge that our expert sample did not capture. Our consistency-oriented setup and findings can speak to the internal validity of our approach, but we leave this kind of evaluation of external validity to future work.

\textbf{Long-term Adaptation.} Our single-session study captured initial reactions to a novel interface. Trust and reliance patterns likely evolve as users develop mental models of system capabilities and limitations. Longitudinal deployment might reveal whether the trust-behavior gap narrows as verification workflows become habitual, or whether users develop stable patterns of selective engagement with provenance features.}

\section{Conclusion}
We present a novel system, \system, that provides argument-grounded provenance information comparing LLM responses in a scholarly QA setting with claims and evidence from source documents. Through a within-subjects user study with 26 researchers, we found that \system significantly reduced participants' trust in LLM outputs compared to standard citation-based provenance. However, this did not translate into different editing behavior change---people edited LLM-generated text similarly across conditions. 
While people valued claim-level verification in principle, time pressure, system latency, and interface complexity prevented meaningful engagement with provenance features. Beyond the interface, our backend architecture for claim-evidence extraction shows promise as an evaluation framework for the trustworthiness of LLM scholarly outputs. The gap between recognizing verification needs and performing verification actions remains a challenge. Our findings indicate that granular provenance information alone is insufficient to change behavior when users face the time pressures and cognitive constraints typical of research settings.







\begin{acks}
We would like to thank our reviewers for their helpful comments. We are grateful to numerous colleagues at NASA including Braxton VanGundy and his team for their guidance on PDF text extraction, and Michael Steele and Charles Liles for their support in deploying PaperTrail, and Shanel Smith for facilitating participant recruitment. We are also grateful to Malik Khadar, Joel Markley, Matthew Zent, and all of the faculty and students in the GroupLens research lab for their feedback and support. We also want to thank everyone who participated in our study.
\end{acks}

\bibliographystyle{ACM-Reference-Format}
\bibliography{000_references}

\appendix
\section{Prompts}\label{app:prompts}

This appendix provides the prompts and JSON schemas used in \system's Argument Extraction Engine. All LLM-based operations use Gemini 2.5 Pro with structured JSON output via the \texttt{response\_json\_schema} parameter. Similarity-based operations (Stage 1 evidence retrieval, Stage 3 evidence verification) do not require prompts.

\subsection{Paper-Level Claim Extraction}\label{app:paper-claim}

\textbf{Pipeline Stage:} Stage 1 (Offline Paper Processing)

\textbf{Purpose:} Extracts atomic scientific claims from individual paragraphs of source documents during offline preprocessing. Applied to each paragraph to build the structured claim database serving as the ground-truth knowledge base.

\textbf{Prompt:}
\begin{quote}
You are an expert research assistant specializing in extracting structured information from scientific texts. Your task is to carefully read the provided scientific paragraph and generate one or more core scientific claims based solely on the information present in that paragraph.

A scientific claim must satisfy the following criteria:
\begin{enumerate}
    \item \textbf{Atomic:} Focus on a single, specific, and indivisible assertion, finding, or conclusion. Avoid compound statements that can be decomposed further.
    \item \textbf{Verifiable:} State something factual whose truthfulness can be checked against evidence or data, either within this paragraph or the broader scientific context.
    \item \textbf{Faithful:} Accurately and precisely reflect the meaning and information given in the source paragraph. Do not introduce outside information or make inferences not directly supported by the text.
    \item \textbf{Decontextualized:} Be understandable as a standalone statement, requiring minimal or no surrounding text from the original paper to grasp its meaning.
    \item \textbf{Declarative:} Be a clear statement or assertion, not a question, hypothesis phrased as a question, or a description of methods or procedures.
\end{enumerate}

Based on these principles, transform the following paragraph into zero or more distinct scientific claims.

\{FEW-SHOT EXAMPLES: 10 paragraph-claims pairs randomly sampled from SciClaimHunt~\cite{kumar2025sciclaimhuntlargedatasetevidencebased}\}

\textbf{Paragraph:} \{PARAGRAPH\}
\end{quote}

\textbf{JSON Schema:}
\begin{verbatim}
{
  "type": "array",
  "items": {
    "type": "object",
    "properties": {
      "claim": {
        "type": "string",
        "description": "A single, atomic claim"
      }
    },
    "required": ["claim"]
  }
}
\end{verbatim}

\textbf{Note:} Evidence retrieval in Stage 1 uses similarity-based extraction with a cosine similarity threshold of 0.75 and does not require an LLM prompt.

\subsection{Answer Generation}\label{app:answer-gen}

\textbf{Pipeline Stage:} Stage 2 (Real-Time Answer Processing)

\textbf{Purpose:} Generates responses to user questions during the scholarly QA session. The answerer LLM receives source documents as context alongside the query, similar to document-grounded question answering where source documents are provided as context. Citation tags enable sentence-level source attribution for the baseline condition and are removed before displaying responses to users.

\textbf{Prompt:}
\begin{quote}
You are an advanced AI research assistant designed to assist users with scholarly literature analysis and question answering. Your primary function is to provide accurate and insightful answers to questions based on one or more scholarly papers provided to you.

The user is performing an editing task and may refer to text they are editing. This text and a description of the editing task will be provided. The conversation history will also be provided if it exists.

In your response, use tags around each sentence to indicate which paper(s) are being referenced. These tags will be removed in post-processing before the answer is shown to the user. Format: \texttt{<Author et al., year> sentence </Author et al., year>}. For sentences drawing on multiple papers, separate citations with semicolons: \texttt{<Author et al., year;Author et al., year>}.

Provide your answer in 300 words or fewer. Do not use formatting such as bullet points or headers.

\textbf{Papers:} \{PAPER\_CONTENTS\}

\textbf{Task description:} \{TASK\_DESCRIPTION\}

\textbf{Text being edited:} \{EDITOR\_TEXT\}

\textbf{Conversation history:} \{CONVERSATION\_HISTORY\}

\textbf{Question:} \{USER\_QUESTION\}
\end{quote}

\subsection{Answer-Level Claim and Evidence Extraction}\label{app:answer-claim}

\textbf{Pipeline Stage:} Stage 2 (Real-Time Answer Processing)

\textbf{Purpose:} Decomposes LLM-generated answers into discrete claims and supporting evidence. Uses the same claim criteria as paper-level extraction to ensure consistency in matching. Extracts both claims and evidence in a single pass since answers are shorter than full papers.

\textbf{Prompt:}
\begin{quote}
You are an expert research assistant specializing in extracting structured information from scientific texts. Your task is to carefully read the provided text and decompose it into discrete claims and their supporting evidence.

A claim must satisfy the following criteria:
\begin{enumerate}
    \item \textbf{Atomic:} Focus on a single, specific, and indivisible assertion. Avoid compound statements that can be decomposed further.
    \item \textbf{Verifiable:} State something factual whose truthfulness can be checked against evidence or data.
    \item \textbf{Faithful:} Accurately reflect the meaning in the source text. Do not introduce outside information.
    \item \textbf{Decontextualized:} Be understandable as a standalone statement.
    \item \textbf{Declarative:} Be a clear statement or assertion, not a question or description of methods.
\end{enumerate}

For each claim identified, extract the exact text spans from the input that express the claim and any text spans that serve as supporting evidence. Text spans must match the original input exactly to enable precise highlighting in the user interface.

\{FEW-SHOT EXAMPLES: text-to-claims-and-evidence pairs demonstrating the expected output format\}

\textbf{Text:} \{ANSWER\_TEXT\}
\end{quote}

\textbf{JSON Schema:}
\begin{verbatim}
{
  "type": "array",
  "items": {
    "type": "object",
    "properties": {
      "claim": {
        "type": "string",
        "description": "A single atomic answer claim"
      },
      "claim_texts": {
        "type": "array",
        "items": {"type": "string"},
        "description": "Verbatim claim texts"
      },
      "evidence_texts": {
        "type": "array",
        "items": {"type": "string"},
        "description": "Verbatim evidence texts"
      }
    },
    "required": ["claim", "claim_texts", 
                 "evidence_texts"]
  }
}
\end{verbatim}

\subsection{Relevant Claims Selection}\label{app:relevant-claims}

\textbf{Pipeline Stage:} Stage 3 (Real-Time Claim Matching)

\textbf{Purpose:} Filters the corpus of pre-extracted paper claims to identify those relevant to the user's question. Operates on candidates pre-filtered by similarity-based retrieval (using cosine similarity between SPECTER embeddings), applying LLM-based semantic reasoning to select the most pertinent claims.

\textbf{Prompt:}
\begin{quote}
You are provided with a set of scientific claims extracted from scholarly papers and a user's question about those papers. Your task is to identify which claims are relevant to answering the question.

A claim is relevant if it:
\begin{itemize}
    \item Directly addresses the question
    \item Provides necessary background information
    \item Contains factual information that would contribute to a complete answer
\end{itemize}

Each claim is accompanied by a numerical ID. Return only the IDs of relevant claims.

\textbf{Question:} \{USER\_QUESTION\}

\textbf{Claims:} \{CLAIM\_LIST\_WITH\_IDS\}
\end{quote}

\textbf{JSON Schema:}
\begin{verbatim}
{
  "type": "array",
  "items": {
    "type": "integer",
    "description": "ID of a relevant claim"
  }
}
\end{verbatim}

\subsection{Relevant Evidence Selection}\label{app:relevant-evidence}

\textbf{Pipeline Stage:} Stage 3 (Real-Time Claim Matching)

\textbf{Purpose:} Selects the most relevant evidence passages for each claim identified in the previous step. For each relevant claim, its supporting evidence forms a sub-corpus that is first filtered by similarity-based retrieval, then the LLM performs final selection of the most pertinent passages.

\textbf{Prompt:}
\begin{quote}
You are provided with a set of evidence passages associated with a scientific claim, along with a user's question. Your task is to identify which evidence passages are most relevant in the context of the question.

Evidence is relevant if it:
\begin{itemize}
    \item Directly supports or substantiates the claim
    \item Provides data, results, or reasoning that validates the claim
    \item Contains contextual information necessary to understand the claim in relation to the question
\end{itemize}

Each evidence passage is accompanied by a numerical ID. Return only the IDs of the most relevant evidence passages.

\textbf{Question:} \{USER\_QUESTION\}

\textbf{Claim:} \{CLAIM\_TEXT\}

\textbf{Evidence passages:} \{EVIDENCE\_LIST\_WITH\_IDS\}
\end{quote}

\textbf{JSON Schema:}
\begin{verbatim}
{
  "type": "array",
  "items": {
    "type": "integer",
    "description": "ID of relevant evidence passage"
  }
}
\end{verbatim}

\subsection{Claim-to-Claim Matching}\label{app:claim-matching}

\textbf{Pipeline Stage:} Stage 3 (Real-Time Claim Matching)

\textbf{Purpose:} Identifies semantic equivalence between claims extracted from the LLM-generated answer and claims from the source papers. The output determines which answer claims are supported by the source literature (displayed as ``Claims included in answer'') and which lack grounding (flagged for user attention).

\textbf{Prompt:}
\begin{quote}
You are provided with two sets of claims: (1) claims extracted from an LLM-generated answer, and (2) claims extracted from source scholarly papers. Your task is to identify which answer claims are semantically equivalent to which paper claims.

Two claims are semantically equivalent if they express the same core assertion, even if worded differently. Minor differences in phrasing, specificity, or elaboration are acceptable provided the fundamental meaning is preserved. Do not match claims that are merely topically related but make different assertions.

For each answer claim that has a match in the paper claims, return the answer claim ID paired with the ID(s) of the matching paper claim(s). Omit answer claims that lack a clear match.

\textbf{Answer claims:} \{ANSWER\_CLAIMS\_WITH\_IDS\}

\textbf{Paper claims:} \{PAPER\_CLAIMS\_WITH\_IDS\}
\end{quote}

\textbf{JSON Schema:}
\begin{verbatim}
{
  "type": "array",
  "items": {
    "type": "object",
    "properties": {
      "answer_claim_id": {
        "type": "integer",
        "description": "ID of the answer claim"
      },
      "paper_claim_ids": {
        "type": "array",
        "items": {"type": "integer"},
        "description": "IDs of semantically equivalent 
                       paper claims"
      }
    },
    "required": ["answer_claim_id", "paper_claim_ids"]
  }
}
\end{verbatim}

\textbf{Note:} Evidence verification in Stage 3 uses cosine similarity with a threshold of $<0.55$ to flag potentially unsupported evidence and does not require an LLM prompt.
\end{document}